%% file: _main.tex
\documentclass[pdflatex,sn-nature]{sn-jnl}

\usepackage{graphicx}%
\usepackage{multirow}%
\usepackage{amsmath,amssymb,amsfonts}%
\usepackage{amsthm}%
\usepackage{mathrsfs}%
\usepackage[title]{appendix}%
\usepackage[table]{xcolor}
\usepackage{textcomp}%
\usepackage{manyfoot}%
\usepackage{booktabs}%
\usepackage{algorithm}%
\usepackage{algorithmicx}%
\usepackage{algpseudocode}%
\usepackage{listings}%
\usepackage{hyperref}       
\usepackage{url}            
\usepackage{subcaption}

\begin{document}

\title{MedForj: An open, large-scale foundational generative prior for high-resolution 3D brain MRI}

\author*[1]{\fnm{Samuel~W.} \sur{Remedios}}\email{samuel.remedios@jhu.edu}
\author[1]{\fnm{Aaron} \sur{Carass}}\email{aaron\_carass@jhu.edu}
\author[1]{\fnm{Jerry~L.} \sur{Prince}}\email{prince@jhu.edu}
\author[2]{\fnm{Blake~E.} \sur{Dewey}}\email{blake.dewey@jhu.edu}
\author[ ]{for the Alzheimer’s Disease Neuroimaging Initiative$^\dagger$}
\author[ ]{for the Australian Imaging Biomarkers and Lifestyle Flagship Study of Aging$^\ddagger$}
\author[ ]{for the Health and Aging Brain Study~(HABS-HD) Study Team$^\S$}
\author[ ]{for the Mayo Clinic Study of Aging$^{\S\S}$}

\affil*[1]{\orgdiv{Image Analysis and Communications Laboratory}, \orgname{Johns Hopkins University}, \orgaddress{\street{3400 North Charles Street}, \city{Baltimore}, \postcode{21218}, \state{MD}, \country{USA}}}
\affil[2]{\orgdiv{Department of Neurology}, \orgname{Johns Hopkins Hospital}, \orgaddress{\street{1800 Orleans St}, \city{Baltimore}, \postcode{21287}, \state{MD}, \country{USA}}}

\abstract{
\unboldmath
This work introduces MedForj, a suite of 3D foundational generative priors based on diffusion models.
The MedForj models were trained on $72{,}659$ 1~mm isotropic 3D $T_1$-weighted MRI human brain image volumes from $38{,}174$ subjects, drawn from a curated corpus of $80{,}675$ volumes from $42{,}506$ subjects spanning $38$ publicly available datasets; two datasets were withheld in their entirety to form an external test set.
These training images were manually inspected to exclude those with poor quality and excessive pathology, and otherwise were minimally processed.
The models include six different diffusion training strategies: rectified flow, latent diffusion rectified flow, flow matching, velocity prediction, clean prediction, and noise prediction.
Image samples produced by each of these models were compared to each other and against real, ground truth data under downstream segmentation distributions, the Fr\'echet Inception Distance~(FID), five inverse problems, and blind human inspection in an observer study.
Flow matching was the strongest strategy overall, achieving the best inverse problem solving results at $28.80$~dB PSNR and $0.874$ SSIM averaged over the five forward problems, the highest rate of reconstructions judged real by blind human raters at $72.6\%$, and the closest per-structure match to real segmented anatomy in a permutation test.
It was not best everywhere: rectified flow produced the most convincing unconditional samples in the observer study~($66.5\%$ judged real versus $57.2\%$) and the best FID~($13.27$ versus $19.34$), and the latent rectified-flow model achieved the smallest joint distributional distance to real anatomy.
No other strategy, however, performed consistently well across all four evaluations.
We therefore recommend flow matching as the default MedForj prior, while releasing every strategy so that the choice can be revisited per application.
All model weights and corresponding code are publicly available at \url{https://github.com/piksl-research/medforj}.
}

\keywords{generative AI, open-source data, diffusion models, neuroimaging}



\maketitle

\section{Introduction}

Many contemporary medical image analysis and processing techniques rely on large, pre-trained foundation models~\cite{foundation_models_survey}.
By learning generalizable representations from diverse datasets, such models can be fine-tuned to downstream tasks such as segmentation~\cite{lee2024foundation}, classification and disease prediction~\cite{van2025foundation} with limited labeled data~\cite{noh2025narrative}, often achieving state-of-the-art performance~\cite{foundation_models_survey,lee2024foundation,van2025foundation}.
Recently, generative models have begun to play a central role in medical image restoration and enhancement, including denoising~\cite{wang2025diffusion}, lesion filling~\cite{zhang2025bi}, bias correction~\cite{lee2023bias,qiu2025memgb}, and super-resolution~\cite{zhao2024mri,friedrich2024wdm,remedios2025cycle}.
In these inverse problems, the pre-trained foundational generative model acts as a probabilistic prior, regularizing solutions to those that are anatomically plausible.
Since image restoration is often the first step of any image analysis pipeline~\cite{ritter2011medical,lauzon2013simultaneous,ades2018evaluation,cai2021prequal}, the availability of general-purpose generative priors that accurately sample realistic image volumes is critical for robust downstream analysis.
Separately, samples drawn from such models can serve as training data for networks solving \emph{other} tasks, such as segmentation, where real labeled data are scarce, expensive to annotate, or restricted from sharing~\cite{fernandez2022can,akbar2024brain,khader2023denoising}.

Despite their promise, training 3D foundational generative models for medical images remains challenging.
The current state of the art in generative modeling leverages diffusion models, a family of techniques that progressively transforms noise into data samples.
The family of diffusion models includes the denoising diffusion probabilistic model~(DDPM)~\cite{ddpm}, the denoising diffusion implicit model~(DDIM)~\cite{DDIM}, score-based models~\cite{chung2022diffusion,chung2022improving}, Schr\"odinger bridges~\cite{chen2022likelihood,shi2023diffusion,su2023dual}, and Brownian bridges~\cite{li2023bbdm,lee2024ebdm,choo2024slice}.
There are also connections between diffusion-based models and flow matching~\cite{liu2023flow,lipman2023flow,gao2025diffusionmeetsflow}.
However, medical image modalities such as magnetic resonance imaging~(MRI) are inherently high-dimensional due to being 3D volumetric data.
This poses a practical problem for diffusion models, which have high computational demands.
Training diffusion models at scale involves balancing constraints related to GPU memory, training time, convergence stability, and the number of image volumes required to adequately capture population-level anatomical variability.

For generative models to function as broadly useful priors in medical imaging, they must satisfy several criteria.
First, they should be trained on large and diverse datasets to capture population-level anatomical variability.
Second, they should generate images of a particular spatial resolution to maintain control over the anatomical detail.
Third, preprocessing should be minimal, ensuring that the learned distribution reflects real-world variability in acquisition protocols and image characteristics.
Finally, trained models and code should be openly available to enable transparent evaluation, reuse, and extension by the community.

At present, no generative models simultaneously meet all of these criteria for 3D brain MRI, though several generative models for brain MRI synthesis have been proposed.
Many approaches operate on 2D slices rather than full 3D volumes~\cite{remedios2023deep,konz2024anatomically,chung2022improving}, at the cost of inter-slice consistency.
Others downsample images to lower spatial resolutions~(e.g., $2$~mm isotropic)~\cite{dorjsembe2022three}, limiting their utility for high-resolution analysis.
Other methods rely on extensive preprocessing such as skull stripping, rigid or nonlinear registration, and bias field correction to simplify the learned data distribution~\cite{dorjsembe2024conditional}, but in doing so implicitly encode assumptions that may not hold across scanners, protocols, or populations.
Latent diffusion models~(LDMs)~\cite{rombach2021high,pinaya2022brain} alleviate memory constraints by operating in a compressed latent space, but LDMs rely on another neural network to perform compression and decompression.
Finally, many existing models are trained on limited datasets or are not released publicly, hindering broader adoption and further research.

To satisfy the aforementioned criteria for generative models in medical image analysis, we introduce MedForj, a suite of 3D generative priors based on diffusion models.
The MedForj models are trained on $72{,}659$ minimally processed, $1$~mm isotropic 3D $T_1$-weighted~($T_1$-w) human brain MRI volumes aggregated from $36$ publicly available datasets, encompassing $38{,}174$ subjects.
All images were manually reviewed by multiple human experts to exclude scans with poor quality or excessive pathology, ensuring a clean and consistent training corpus while avoiding aggressive algorithmic preprocessing.
In this paper, we compare six diffusion model training strategies: noise prediction, clean prediction, velocity prediction, flow matching, rectified flow, and latent diffusion rectified flow.
To evaluate these strategies' efficacy towards building a useful generative prior, we compare them across inverse problems, anatomical segmentation accuracy at a distributional level, an observer study, and Frech\'et Inception Distance~(FID)~\cite{heusel2017gans,Seitzer2020FID}.
MedForj is the first broadly available, high-resolution, 3D generative model suite for the generation of synthetic $T_1$-w brain MRI.

We contribute (i) a controlled comparison of six diffusion training strategies for 3D brain MRI trained on the same 72,659-volume corpus with the same U-Net backbone, optimizer, schedule, and augmentation, and sampled at the same operating point, so that the training objective is the only intended difference among them; (ii) an evaluation of each trained model as a plug-in prior, scored against ground truth on five MRI inverse problems and assessed further by FID, segmented-anatomy distributions, and a blind observer study; and (iii) the public release of all six sets of trained weights together with the training and inference code.

\newpage
\section{Results}
\label{sec:results}
We tested all six training strategies used in developing the MedForj suite.
The following results include four primary experimental evaluations~(inverse problems, FID, downstream segmentation distributions, blind observer study).
This paper uses a diverse dataset spanning $38$ open-source dataset \textit{sources} with a total of $80,675$ 1~mm isotropic 3D $T_1$-weighted MRI human brain image volumes.
This dataset was partitioned into three categories: \textit{internal training}, \textit{internal testing}, and \textit{external testing}.
The internal training set includes $72,659$ manually curated $T_1$-w MR image from subjects selected from $36$ of the sources.
The internal testing set includes subjects withheld from training from those same $36$ sources.
The external testing set includes all subjects from the remaining $2$ sources, entirely withheld from training.
Those two external testing sources were designated before any training and were not selected for any distributional property.
A summary of the data split by source is provided in Table~\ref{tab:datasets}.

First, we show how the investigated strategies compare across five inverse problem tasks in MRI, evaluating their utility as priors.
Second, we show the generated images themselves and compare their distributions with FID.
Third, we compare the distributions of their segmentations compared to non-generated images, evaluating their utility as generative models.
Fourth, we show the outcome of a human rater observer study, evaluating their realism for both inverse-problem-solved images and generated images.
Throughout these results, all metrics are reported on the internal and external testing sets, in total comprising $8,016$ image volumes from the $38$ dataset sources.
We defer a full discussion of the training strategies, dataset curation, and other dataset details to Section~\ref{sec:methods}.

\begin{table}[ht]
\centering
\caption{Distribution of subjects~(S) and image volumes~(V) by dataset. Note that this is not the total number of data available per dataset, but rather the amount of data after the exclusion described in Sec.~\ref{sec:datasets} and Sec.~\ref{sec:curation}.}
\begin{tabular}{lrrrrrr}
\toprule
\textbf{Dataset Name} & \textbf{Subjects~(S)} & \textbf{Volumes~(V)} & \textbf{Train S} & \textbf{Train V} & \textbf{Test S} & \textbf{Test V} \\
\midrule
A4 & 1742 & 6903 & 1594 & 6320 & 148 & 583 \\
ABIDE & 781 & 781 & 715 & 715 & 66 & 66 \\
ABIDEII & 959 & 1221 & 878 & 1112 & 81 & 109 \\
ADNI & 2587 & 17965 & 2368 & 16505 & 219 & 1460 \\
AIBL~\cite{dataset-aibl} & 673 & 1203 & \textemdash & \textemdash & 673 & 1203 \\
AOMIC-ID1000~\cite{dataset-aomic} & 928 & 2768 & 850 & 2539 & 78 & 229 \\
AOMIC-PIOP1~\cite{dataset-aomic} & 216 & 216 & 198 & 198 & 18 & 18 \\
AOMIC-PIOP2~\cite{dataset-aomic} & 226 & 226 & 207 & 207 & 19 & 19 \\
BHRC~\cite{dataset-bhrc} & 609 & 902 & 558 & 825 & 51 & 77 \\
CCNP~\cite{dataset-ccnp} & 195 & 195 & 179 & 179 & 16 & 16 \\
COBRE~\cite{dataset-cobre} & 193 & 1275 & 177 & 1165 & 16 & 110 \\
COGTRAIN~\cite{dataset-cogtrain} & 166 & 293 & 152 & 271 & 14 & 22 \\
CORR~\cite{dataset-corr} & 1178 & 2332 & 1078 & 2113 & 100 & 219 \\
DLBS~\cite{dataset-dlbs} & 314 & 314 & 288 & 288 & 26 & 26 \\
FCON1000~\cite{dataset-fcon1000} & 563 & 563 & 516 & 516 & 47 & 47 \\
GSP~\cite{dataset-gsp} & 1570 & 1639 & 1437 & 1502 & 133 & 137 \\
HABS-HD & 4233 & 6437 & 3874 & 5880 & 359 & 557 \\
HBN~\cite{dataset-hbn} & 2398 & 2398 & 2195 & 2195 & 203 & 203 \\
HCP1200 & 1112 & 1112 & 1018 & 1018 & 94 & 94 \\
ICBM & 444 & 444 & 407 & 407 & 37 & 37 \\
IXI & 581 & 581 & 532 & 532 & 49 & 49 \\
MCSA & 1799 & 3087 & 1647 & 2841 & 152 & 246 \\
MPILEMON~\cite{dataset-mpilemon} & 226 & 226 & 207 & 207 & 19 & 19 \\
NACC & 3680 & 5472 & 3368 & 4994 & 312 & 478 \\
NADR~\cite{dataset-nadr} & 301 & 301 & 276 & 276 & 25 & 25 \\
NARRATIVES~\cite{dataset-narratives} & 315 & 348 & 289 & 322 & 26 & 26 \\
NEUROPHENOM~\cite{dataset-neurophenom} & 265 & 265 & 243 & 243 & 22 & 22 \\
NIMHHRVD~\cite{dataset-nihmhrvd} & 249 & 499 & 228 & 456 & 21 & 43 \\
NKI~\cite{dataset-nkirockland} & 1314 & 2254 & 1203 & 2058 & 111 & 196 \\
OASIS3~\cite{dataset-oasis3} & 1340 & 3493 & 1227 & 3204 & 113 & 289 \\
OASIS4~\cite{dataset-oasis4} & 655 & 723 & 600 & 663 & 55 & 60 \\
PNC~\cite{dataset-pnc} & 1600 & 1600 & 1464 & 1464 & 136 & 136 \\
PPMI & 1975 & 3748 & 1808 & 3433 & 167 & 315 \\
QTIM~\cite{dataset-qtim} & 1199 & 1339 & 1098 & 1222 & 101 & 117 \\
SALD~\cite{dataset-sald} & 494 & 494 & 453 & 453 & 41 & 41 \\
SCAN & 4702 & 5886 & 4303 & 5384 & 399 & 502 \\
SLEEP~\cite{dataset-sleep} & 136 & 136 & \textemdash & \textemdash & 136 & 136 \\
SLIM~\cite{dataset-slim} & 588 & 1036 & 539 & 952 & 49 & 84 \\
\midrule
Total & 42506 & 80675 & 38174 & 72659 & 4332 & 8016 \\
\bottomrule
\end{tabular}
\label{tab:datasets}
\end{table}

\subsection{Inverse Problem Results}
We compared each training strategy under five inverse problems appropriate for MRI: through-plane super-resolution, motion correction, k-space acceleration, inpainting, and Rician denoising.
Full details of the form of each forward problem, as well as the diffusion posterior sampling~(DPS) and ReSample parameters, are described in Section~\ref{sec:methods}.

\begin{figure}
    \centering 
    \includegraphics[width=\linewidth]{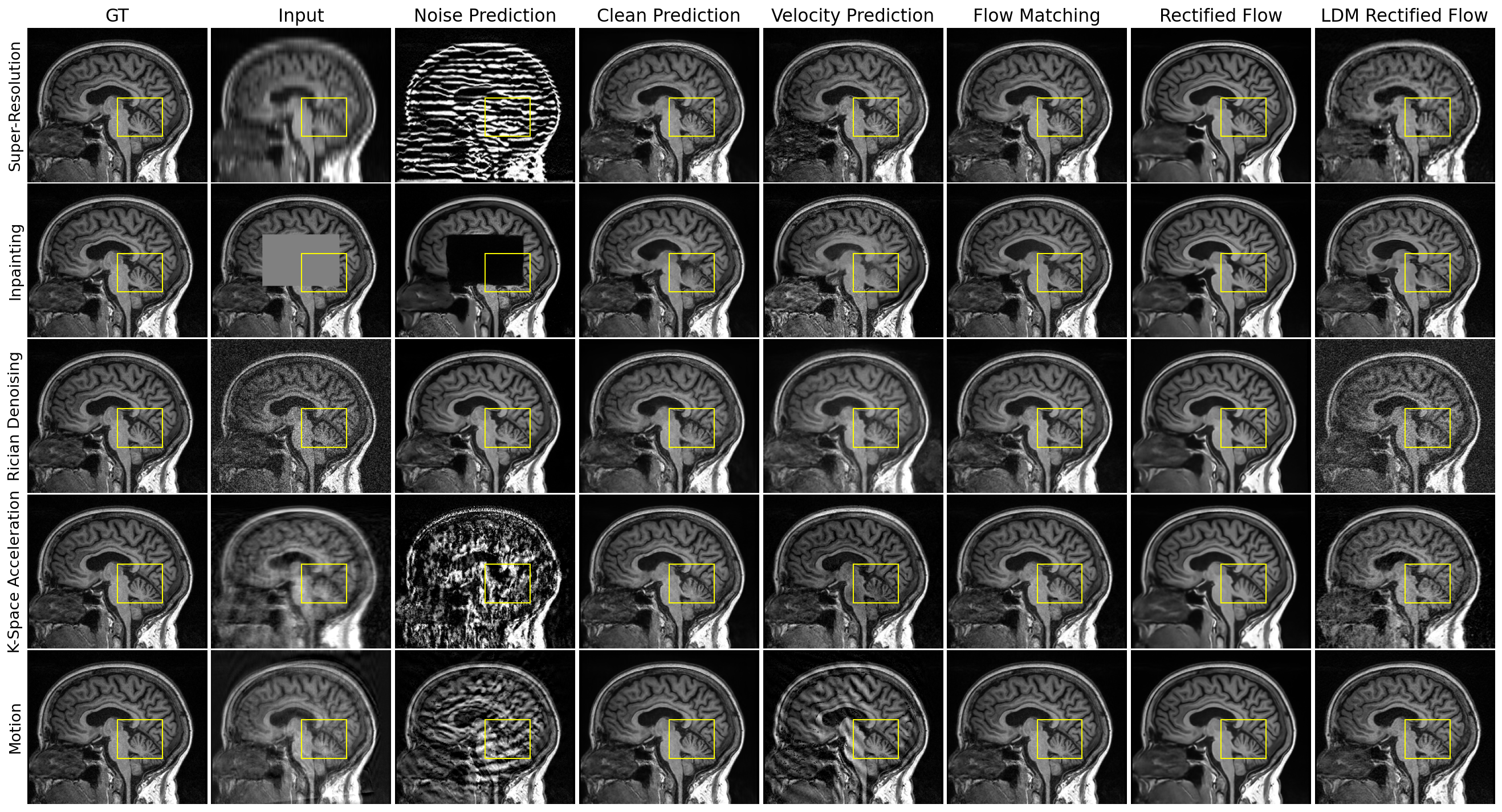}
    \caption{A sagittal slice of a representative image used in the inverse problem experiments, with each row corresponding to a different inverse problem. The first column is identical across all rows, showing the ground truth. The second column shows the degraded input for each method. Following columns show the result of each strategy for solving the inverse problem. Yellow boxes indicate a zoomed region, shown in Fig.~\ref{fig:inv_probs_zoom}.}
    \label{fig:inv_probs_full}
\end{figure}

\begin{figure}
    \centering 
    \includegraphics[width=\linewidth]{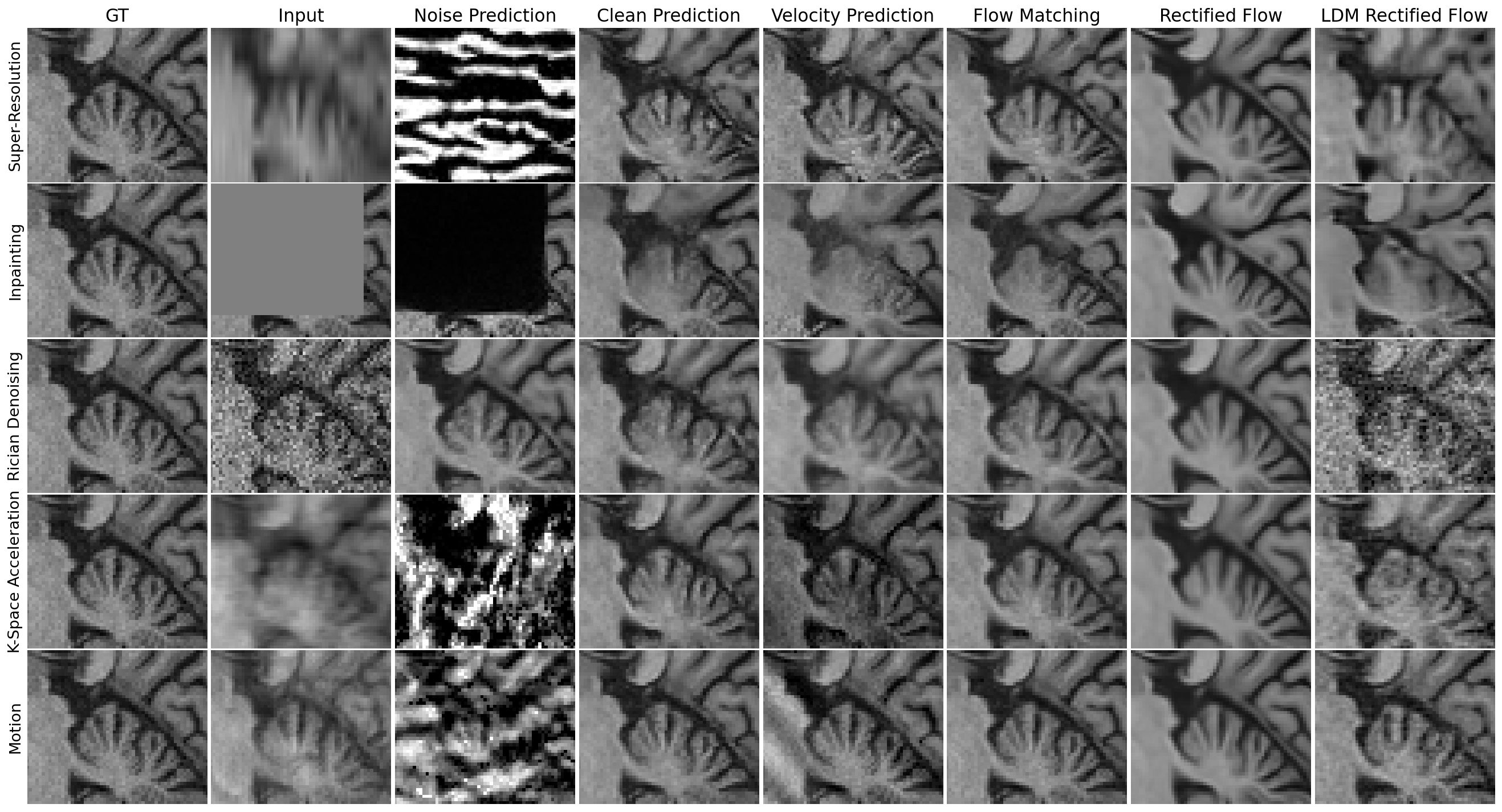}
    \caption{Zoomed regions corresponding to the yellow boxes in Fig.~\ref{fig:inv_probs_full}. Rows and columns follow the same layout: one inverse problem per row, ground truth and degraded input in the first two columns, and one training strategy per remaining column.}
    \label{fig:inv_probs_zoom}
\end{figure}

A representative qualitative result is shown in Figure~\ref{fig:inv_probs_full}, with zoomed insets displayed separately in Figure~\ref{fig:inv_probs_zoom}.
Each row is a different inverse problem to solve, and each column is the training strategy used.
The column labeled ``GT'' is the ground truth high-resolution image.
The column labeled ``Input'' shows the input for the inverse problem solving technique, except for super-resolution, where the input image is displayed in the figure using linear interpolation.
The noise prediction strategy failed to produce reliable volumes in nearly every case, due to the model failing to converge~(its unconditional samples are shown in Fig.~\ref{fig:gen-results}).
Aside from noise prediction, most other results achieve plausible results. 
Of the non-noise prediction strategies, clear apparent errors occur for velocity prediction on Rician denoising and motion, and LDM rectified flow on super-resolution, Rician denoising, and k-space acceleration.
Regarding the zoomed region in Fig.~\ref{fig:inv_probs_zoom}, clean prediction, flow, and rectified flow achieve the most consistent results, particularly at the superior portion of the cerebellum. 

Quantitative results are reported in Table~\ref{tab:inverse}.
Flow matching overall performs best across peak signal-to-noise ratio~(PSNR), structural similarity index measure~(SSIM)~\cite{wang2004image}, and learned perceptual image patch similarity~(LPIPS)~\cite{zhang2018unreasonable} metrics.
It is only outperformed by rectified flow in PSNR and SSIM on super-resolution and Rician denoising, in SSIM by clean prediction on k-space acceleration, in LPIPS by velocity prediction on super-resolution and motion correction, and in LPIPS by noise prediction on Rician denoising.
When considering the average score of all inverse problems, flow matching significantly outperforms all other methods.
A description of the statistical test is provided in the caption of Table~\ref{tab:inverse}.

\begin{table}[!ht]
\centering
\small
\caption{Quantitative results (PSNR, SSIM, LPIPS) for the inverse problem solving task over $100$ subjects. The final row for each metric shows the average score across all tasks for that metric. ``Input'' is the input image for each inverse problem, except for super-resolution whose images are linearly interpolated. Other abbreviations: ``Vel.'' is velocity prediction, ``R. Flow'' is rectified flow, and ``LDM'' is the latent diffusion model rectified flow. \textbf{Bolded} numbers indicate the best scoring strategy for that task, and asterisks$^{*}$ indicate statistical significance~($\alpha = 0.05$) to a two-sided paired Wilcoxon signed-rank test~\cite{wilcoxon1945individual} after Holm-Bonferroni~\cite{holm1979simple} correction. }
\begin{tabular}{lrrrrrrr}
\toprule
\textbf{Forward Problem} & \textbf{Input} & \textbf{Noise} & \textbf{Clean} & \textbf{Vel.} & \textbf{Flow} & \textbf{R.\ Flow} & \textbf{LDM} \\
\midrule
\multicolumn{8}{l}{\textit{PSNR (dB) $\uparrow$}} \\
Super-resolution & 21.95 & 13.47 & 24.53 & 24.02 & 24.77 & \textbf{25.17}$^{*}$ & 23.09 \\
Inpainting & 26.83 & 20.73 & 31.64 & 26.07 & \textbf{32.76}$^{*}$ & 28.06 & 25.11 \\
Rician denoising & 18.90 & 25.53 & 25.68 & 21.38 & 25.24 & \textbf{26.97}$^{*}$ & 17.76 \\
$k$-space accel. & 19.25 & 16.36 & 27.91 & 24.87 & \textbf{28.10}$^{*}$ & 27.03 & 22.88 \\
Motion & 21.76 & 21.31 & 31.97 & 25.40 & \textbf{33.14}$^{*}$ & 29.84 & 25.82 \\
\cmidrule(lr){2-8}
\textit{All inverse problems} & 21.74 & 19.48 & 28.35 & 24.35 & \textbf{28.80}$^{*}$ & 27.41 & 22.93 \\
\midrule
\multicolumn{8}{l}{\textit{SSIM $\uparrow$}} \\
Super-resolution & 0.718 & 0.508 & 0.823 & 0.807 & 0.824 & \textbf{0.837}$^{*}$ & 0.751 \\
Inpainting & 0.945 & 0.847 & 0.928 & 0.882 & \textbf{0.954}$^{*}$ & 0.869 & 0.800 \\
Rician denoising & 0.414 & 0.823 & 0.815 & 0.652 & 0.795 & \textbf{0.839}$^{*}$ & 0.340 \\
$k$-space accel. & 0.466 & 0.511 & \textbf{0.871}$^{*}$ & 0.796 & 0.862 & 0.853 & 0.606 \\
Motion & 0.748 & 0.789 & 0.921 & 0.880 & \textbf{0.932}$^{*}$ & 0.892 & 0.810 \\
\cmidrule(lr){2-8}
\textit{All inverse problems} & 0.658 & 0.695 & 0.872 & 0.803 & \textbf{0.874}$^{*}$ & 0.858 & 0.661 \\
\midrule
\multicolumn{8}{l}{\textit{LPIPS $\downarrow$}} \\
Super-resolution & 0.335 & 0.377 & 0.103 & \textbf{0.071}$^{*}$ & 0.083 & 0.173 & 0.149 \\
Inpainting & 0.081 & 0.161 & 0.052 & 0.108 & \textbf{0.026}$^{*}$ & 0.129 & 0.063 \\
Rician denoising & 0.428 & \textbf{0.084}$^{*}$ & 0.122 & 0.155 & 0.099 & 0.149 & 0.432 \\
$k$-space accel. & 0.358 & 0.387 & 0.076 & \textbf{0.049}$^{*}$ & 0.052 & 0.163 & 0.183 \\
Motion & 0.106 & 0.152 & 0.053 & 0.083 & \textbf{0.028}$^{*}$ & 0.119 & 0.048 \\
\cmidrule(lr){2-8}
\textit{All inverse problems} & 0.262 & 0.232 & 0.081 & 0.093 & \textbf{0.058}$^{*}$ & 0.147 & 0.175 \\
\bottomrule
\end{tabular}
\label{tab:inverse}
\end{table}

\textbf{The autoencoder ceiling:}\quad
A LDM cannot reconstruct an image more faithfully than its autoencoder can reproduce that image, however well the diffusion process itself is solved.
We measured this limit directly for the pretrained MAISI autoencoder~\citep{guo2025maisi,zhao2026maisi} that our latent model uses: each clean ground-truth volume was encoded and decoded, and the decoded volume was scored against the original.
On the same $100$ held-out subjects from the inverse problem task, this autoencoding yielded a mean $26.66$~dB PSNR and $0.850$ SSIM, providing an empirical distortion metric ceiling for LDM's performance on inverse problem solving.
Conversely, the best non-LDM prior surpassed this ceiling on three of the five forward problems: Rician denoising~($26.97$~dB), $k$-space acceleration~($28.10$~dB), and motion correction~($33.14$~dB).
The best non-LDM prior also exceeds the ceiling on inpainting~($32.76$~dB), because whole-volume PSNR for inpainting is dominated by the large untouched periphery rather than by the recovered region.
Only slice-selection super-resolution remains below the ceiling for all priors.
Thus, part of the gap between the LDM and non-LDM priors in Table~\ref{tab:inverse} is architectural due to the limits of the autoencoder.
We note that this is not a failing of the particular autoencoder for MAISI; autoencoders do not achieve perfect reconstruction for arbitrary inputs in their domain, and therefore this limitation will exist for any LDM-based generative prior.

\subsection{Generative Results}
\begin{figure}
    \centering 
    \includegraphics[width=\linewidth]{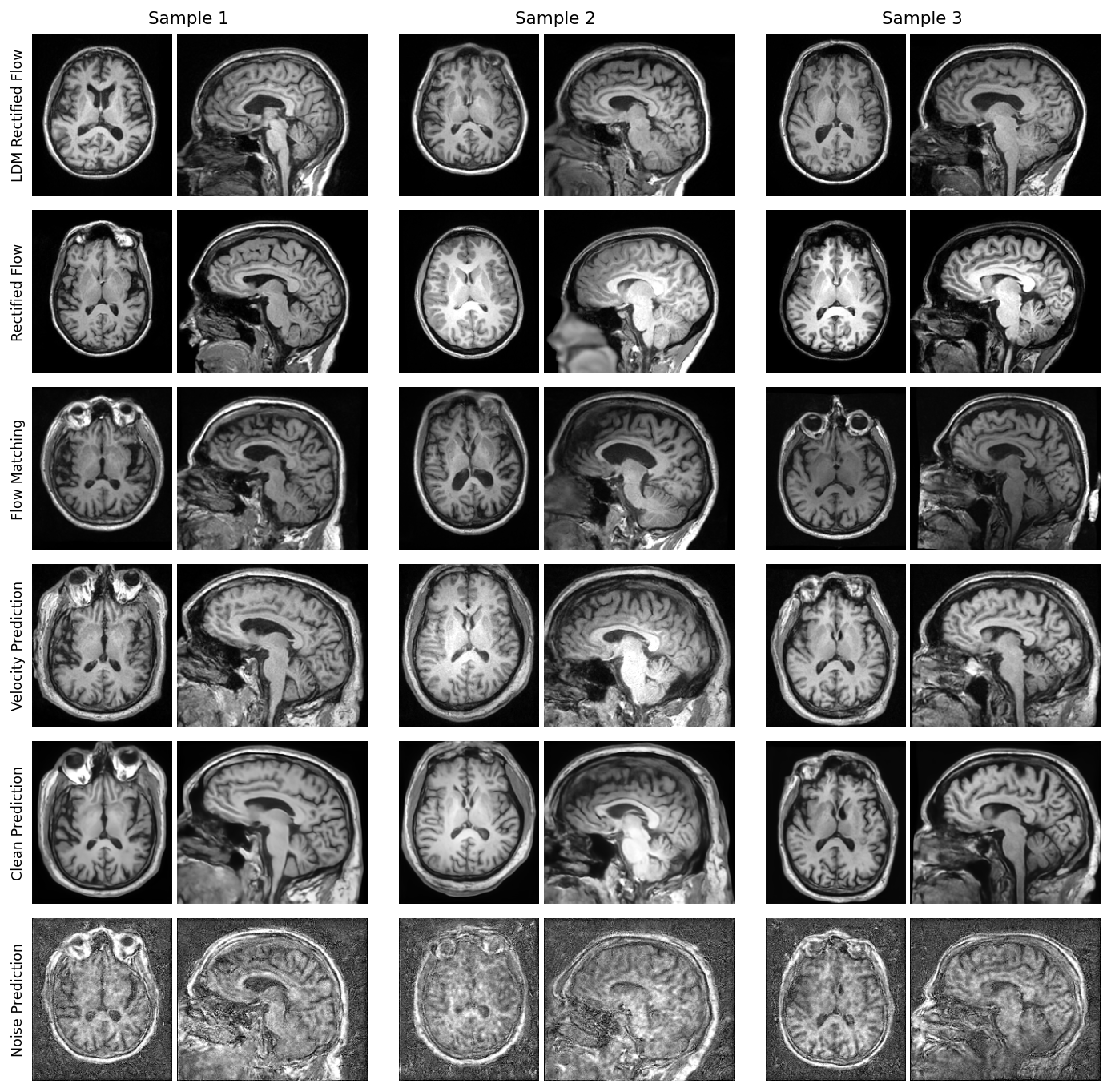}
    \caption{Biplanar views of three uncurated samples from each training strategy. These three samples share the exact random seed across all training strategies. All volumes were generated with 256 DDIM steps.}
    \label{fig:gen-results}
\end{figure}
In Fig.~\ref{fig:gen-results}, three uncurated samples are shown for each training strategy.
Biplanar views are rendered to demonstrate the 3D volumetric consistency.
Aside from the noise prediction results, where the model failed to converge, remaining methods produced feasible high-resolution images with variability in contrast, anatomy, and head position and size.
Clean prediction and velocity prediction appear to have some implausible whole-head anatomies.
Flow, rectified flow, and LDM rectified flow models exhibit more natural head shapes.
The third sample for flow matching appears to have generated an image with wraparound, though clearly no exact wraparound has occurred. 

\begin{figure}
    \centering 
    \includegraphics[width=\linewidth]{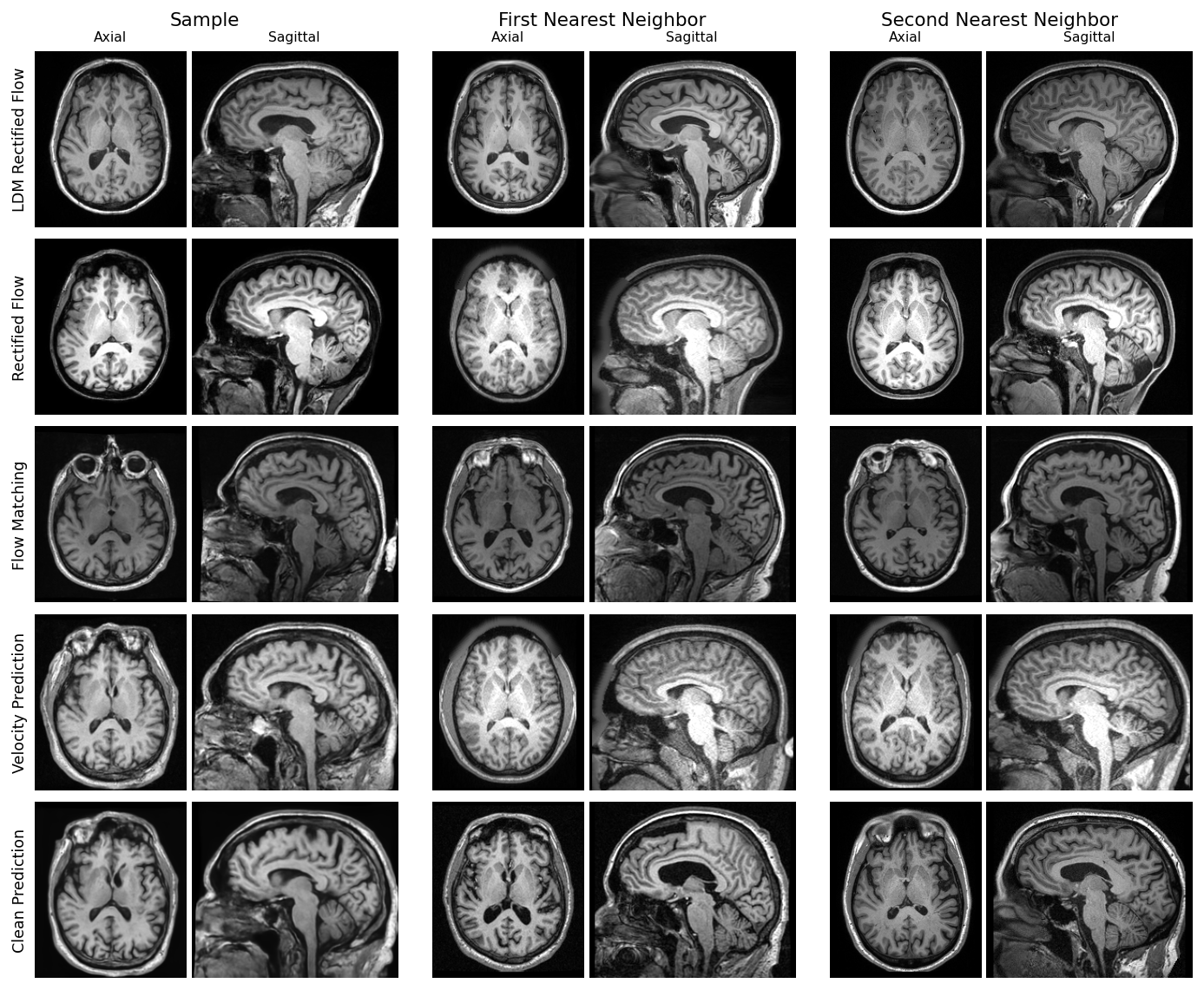}
    \caption{Biplanar views of synthetic images and their two nearest neighbors.
    Nearest neighbor is computed as the minimum mean squared error in voxel space.
    }
    \label{fig:nearest-neighbors}
\end{figure}

In Fig.~\ref{fig:nearest-neighbors}, we show biplanar views of samples from each training strategy alongside their two nearest neighbors in the training dataset, demonstrating that while the generative models can generate coherent 3D images, they did not memorize the dataset.

To show the range of sample quality within each strategy rather than a single representative draw, the supplementary material additionally reports the best- and worst-scoring sample from every strategy, ranked by the distance of each sample's image features from those of the real training data.

Images were generated with $256$ DDIM~\cite{DDIM} steps, i.e.\ $256$ number of function evaluations~(NFE).
This number was found through an ablation study over the number of steps, comparing FID~\cite{heusel2017gans,Seitzer2020FID} to the internal training set.
The ablation study was designed to find the best ``fair match'' across all training strategies.
Here, we defined a fair match to be the fewest number of denoising steps within 10\% of each strategy's best-performing number of denoising steps, shared across all strategies.
The result $256$ was the sole winner of this approach.
The full results of this ablation study are shown in Table~\ref{tab:ablation-steps}.

\begin{table}[!ht]
\centering
\small
\caption{Ablation study to select the number of denoising steps, using FID between $100$ samples from each strategy and the entire internal training corpus. \textbf{Bolded} cells give each strategy's best~(lowest) FID. Daggered$^{\dagger}$ cells exceed that strategy's own best FID by more than $10\%$ and so fail the selection rule at that step count. The operating point is the fewest steps at which \emph{every} strategy is within $10\%$ of its own best FID; this is $256$, identically against all three real references. Noise prediction$^{\ddagger}$ is reported for completeness but excluded from the rule, as it failed to converge and its FID never approaches that of the other strategies. A stricter $5\%$ bound was pre-registered first and returned no step count common to all strategies, so the $10\%$ bound reported here is the fallback specified by the same pre-registration rather than a threshold chosen after seeing the data.}
\begin{tabular}{lrrrrrrr}
\toprule
& \multicolumn{7}{c}{\textbf{Denoising steps (NFE)}} \\
\cmidrule(lr){2-8}
\textbf{Strategy} & 16 & 32 & 64 & 128 & \textbf{256} & 512 & 1000 \\
\midrule
Noise ($\epsilon$)$^{\ddagger}$ & \textbf{131.91} & 135.34 & 141.09 & 144.74 & 142.12 & 140.54 & 143.70 \\
Clean ($x_0$) & 58.07$^{\dagger}$ & 55.17 & \textbf{52.24} & 54.63 & 56.76 & 58.99$^{\dagger}$ & 58.40$^{\dagger}$ \\
Velocity & 47.23$^{\dagger}$ & 42.41$^{\dagger}$ & 37.43 & 37.46 & \textbf{35.99} & 37.19 & 37.10 \\
Flow & 36.74$^{\dagger}$ & 29.95$^{\dagger}$ & 23.44$^{\dagger}$ & 22.54$^{\dagger}$ & 21.50 & 20.86 & \textbf{20.43} \\
Rectified flow & 59.51$^{\dagger}$ & 34.55$^{\dagger}$ & 23.36$^{\dagger}$ & 16.41$^{\dagger}$ & 15.52 & \textbf{14.27} & 14.38 \\
LDM (rect.\ flow) & 36.84$^{\dagger}$ & 19.32$^{\dagger}$ & 17.13$^{\dagger}$ & 16.74$^{\dagger}$ & 15.57 & 15.40 & \textbf{14.81} \\
\bottomrule
\end{tabular}
\label{tab:ablation-steps}
\end{table}

\newpage
\subsection{Distributional Realism}

We perform three comparisons of generated samples to evaluate the strategies across different axes.
First, we compare their feature and intensity distributions with the Fr\'echet Inception Distance~(FID)~\cite{heusel2017gans,Seitzer2020FID}.
Then, we perform anatomical segmentation with SynthSeg~\cite{billot_synthseg_2023} and compare six regions of interest~(cerebral cortex, brain stem, ventricles, thalamus, putamen, cerebellar cortex).
With those segmentations, we characterize the volume distributions of those regions of interest and report mean and variance drift compared to real data.
We then perform a per-structure bootstrapped Kolmogorov--Smirnov permutation test to evaluate whether the volumes of structures are statistically distinguishable from real volumes.
Third, we use the WASABI metric~\cite{jafrasteh2025wasabi} to perform joint distribution comparison.

\subsubsection{Feature and intensity distribution with FID}
\label{sec:fid}
The Fr\'echet Inception Distance~(FID)~\cite{heusel2017gans,Seitzer2020FID} was used to compare the quantitative realism of the generated image volumes of each training strategy.
The results for FID are shown in Table~\ref{tab:fid}; a lower FID score is better.
The columns labeled \textbf{Train~(Int.)}, \textbf{Test~(Int.)}, and \textbf{Test~(Ext.)} show the FID between the pairs of real datasets, serving as a reference.

We include the FID between real data splits as an interpretable reference when evaluating the FID between synthetic and real data.
We do this because the FID computed between medical images differs from conventional values in natural images that expect FID below $4$~\cite{rombach2021high,jit}.
The FID between internal training and testing was $0.09$; between internal training and external was $9.66$; and between internal testing and external datasets was $9.93$.
This provides an acceptable real--real FID value of approximately $9.9$ between two datasets of $T_1$-w images for 2D slices of volumetric MRI. 
The last six columns of Table~\ref{tab:fid} show the FID between real and synthetic datasets.
Flow matching, rectified flow, and the latent rectified-flow model achieve substantially better FID than the other three strategies, and rectified flow reaches $13.27$.

\begin{sidewaystable}[!ht]
\centering
\setlength{\tabcolsep}{3pt}
\caption{Fr\'echet Inception Distance (FID) between pairs of datasets of 2D slices extracted from each volume. A lower FID is better. Internal is denoted ``(Int.)'' and external ``(Ext.)''. Each row denotes the real dataset being compared against. The columns labeled ``Train (Int.)'', ``Test (Int.)'' and ``Test (Ext.)'' report FID between real data splits, establishing the range of FID scores that are acceptable for synthetic images. The remaining columns report FID between real data and samples from each training strategy.}
\begin{tabular}{lrrrrrrrrr}
\toprule
\textbf{Ref.} & \textbf{Train (Int.)} & \textbf{Test (Int.)} & \textbf{Test (Ext.)} & \textbf{Noise} & \textbf{Clean} & \textbf{Velocity} & \textbf{Flow} & \textbf{R.\ Flow} & \textbf{LDM} \\
\midrule
Train (Int.) & --- & 0.09 & 9.66 & 140.74 & 54.57 & 34.80 & 19.34 & 13.27 & 13.40 \\
Test (Int.) & 0.09 & --- & 9.93 & 141.25 & 54.87 & 34.82 & 19.22 & 13.59 & 13.57 \\
Test (Ext.) & 9.66 & 9.93 & --- & 138.98 & 69.56 & 44.45 & 29.02 & 28.27 & 26.46 \\
\bottomrule
\end{tabular}
\label{tab:fid}
\end{sidewaystable}

\subsubsection{Joint distributional distance in segmented anatomy}
We use the WASABI metric~\cite{jafrasteh2025wasabi} to evaluate generative realism in segmentation volumes.
In similar spirit to the empirical real--real FID floor described in Sec.~\ref{sec:fid}, we compute WASABI between segmentations of two randomly sampled halves of the external test set.
Each strategy compares to the segmentations from the full external test set. 
These results are shown in Table~\ref{tab:wasabi}.
The latent rectified-flow model was closest to real at $3.47$, followed by flow matching at $4.20$.
Even the closest cohort, however, sits at $6.4$ times the real--real WASABI floor.

\begin{table}[!ht]
\centering
\caption{The WASABI score for each training strategy. \textbf{SD} is the standard deviation of the WASABI score. \textbf{$\times$floor} is the row's distance divided by the \emph{real vs.\ real} distance in the first row.}
\begin{tabular}{lrrr}
\toprule
\textbf{Strategy} & \textbf{WASABI} ($\times 10^{4}$) $\downarrow$ & \textbf{SD} & \textbf{$\times$floor} \\
\midrule
\textit{Real vs.\ real} & \textit{0.54} & \textit{0.12} & --- \\
\midrule
Noise ($\epsilon$) & 32.77 & 5.14 & 60.7 \\
Clean ($x_0$) & 6.09 & 0.82 & 11.3 \\
Velocity & 5.70 & 0.71 & 10.6 \\
Flow & 4.20 & 0.78 & 7.8 \\
Rectified flow & 9.99 & 1.25 & 18.5 \\
LDM (rect.\ flow) & \textbf{3.47} & 0.52 & 6.4 \\
\bottomrule
\end{tabular}
\label{tab:wasabi}
\end{table}

\subsubsection{Comparison of segmentation distributions with permutation testing}
\begin{figure}
    \centering 
    \includegraphics[width=\linewidth]{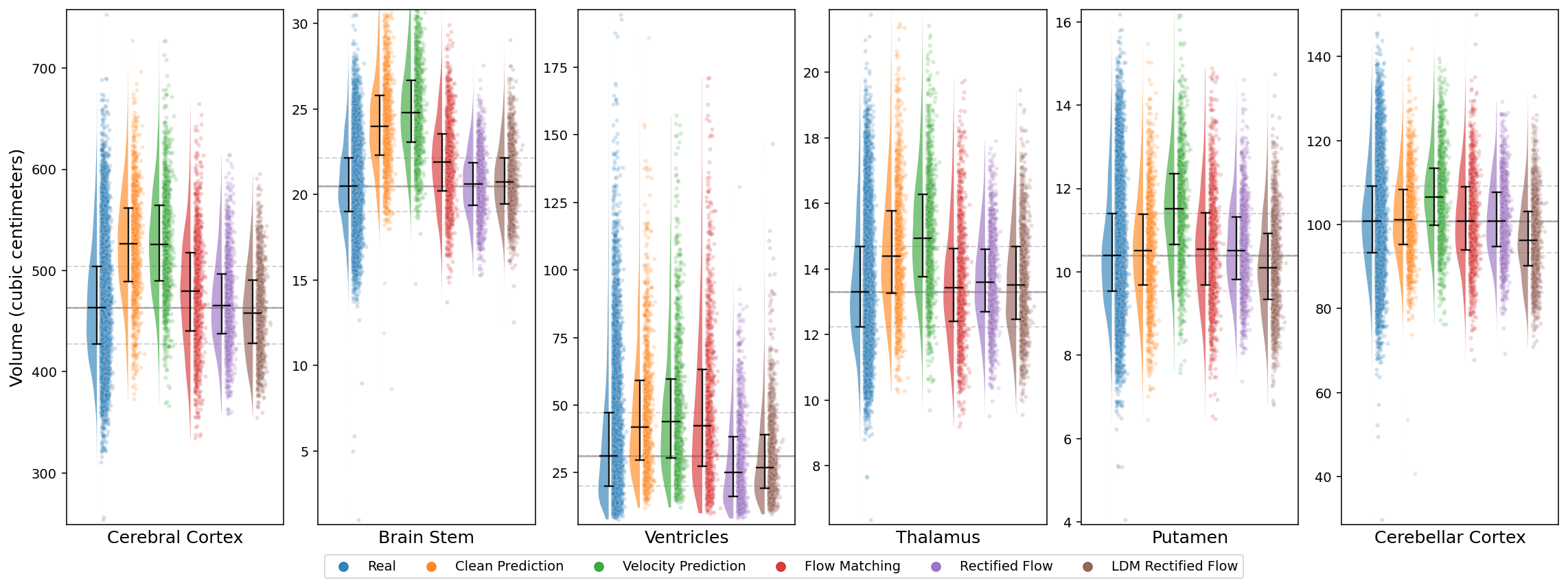}
    \caption{Raincloud plots showing the distributions of volumes from six regions of interest as computed by SynthSeg. Horizontal lines extend from the real distribution quartiles to aid comparison across strategies.}
    \label{fig:raincloud}
\end{figure}

To determine whether the synthetic image volumes are anatomically plausible, SynthSeg~\cite{billot_synthseg_2023} was applied to all results and distributions of six regions of interest~(cerebral cortex, brain stem, ventricles, thalamus, putamen, cerebellar cortex) were compared.
The segmented volume distribution of these six regions are shown in Fig.~\ref{fig:raincloud}.
While the distributions qualitatively appear close, there is apparent mean and variance shift between the real data and the synthetic data.

We first characterize how the synthetic distributions differ from real, then test whether the difference is statistically significant.
In Table~\ref{tab:drift}, we report the percentage drift in mean and standard deviation in segmented structure volume of each training strategy from real data.

\begin{table}[!ht]
\centering
\setlength{\tabcolsep}{4pt}
\caption{Per-structure drift of the synthetic SynthSeg volume distributions from real, for all six strategies. \textbf{Mean drift} is $100\times(\bar{x}_{\mathrm{model}} - \bar{x}_{\mathrm{real}}) / \bar{x}_{\mathrm{real}}$, so positive means the synthetic structure is larger than real. \textbf{Cohen's $d$} restates that difference in pooled-SD units so structures of very different absolute size can be compared. \textbf{SD ratio} is the model's standard deviation divided by the real standard deviation for that structure: $1.0$ reproduces the real spread, below $1.0$ means the model generates less anatomical volume variation than real, above $1.0$ means more.}
\begin{tabular}{lrrrrrr}
\toprule
\textbf{Structure} & Noise & Clean & Velocity & Flow & Rectified flow & LDM \\
\midrule
\multicolumn{7}{l}{\textit{Mean drift (\%) --- positive is larger than real}} \\
Cerebral cortex & $-2.2$ & $+11.6$ & $+12.4$ & $+2.0$ & $-0.5$ & $-1.9$ \\
Brain stem & $-33.2$ & $+15.9$ & $+20.3$ & $+5.7$ & $-0.4$ & $+1.1$ \\
Ventricles & $-13.6$ & $+28.8$ & $+30.8$ & $+29.3$ & $-21.6$ & $-14.9$ \\
Thalamus & $-44.3$ & $+6.9$ & $+11.1$ & $+0.2$ & $+0.4$ & $+0.1$ \\
Putamen & $-48.9$ & $-0.7$ & $+8.9$ & $+0.1$ & $+0.1$ & $-4.1$ \\
Cerebellar cortex & $-42.1$ & $-0.2$ & $+5.2$ & $-0.3$ & $-0.7$ & $-4.9$ \\
\addlinespace\midrule
\multicolumn{7}{l}{\textit{Cohen's $d$ --- drift in pooled-SD units}} \\
Cerebral cortex & $-0.15$ & $+0.98$ & $+1.05$ & $+0.17$ & $-0.04$ & $-0.17$ \\
Brain stem & $-1.28$ & $+1.28$ & $+1.62$ & $+0.48$ & $-0.04$ & $+0.10$ \\
Ventricles & $-0.22$ & $+0.45$ & $+0.47$ & $+0.42$ & $-0.39$ & $-0.27$ \\
Thalamus & $-1.73$ & $+0.50$ & $+0.81$ & $+0.01$ & $+0.03$ & $+0.01$ \\
Putamen & $-1.87$ & $-0.06$ & $+0.68$ & $+0.01$ & $+0.01$ & $-0.33$ \\
Cerebellar cortex & $-1.64$ & $-0.02$ & $+0.46$ & $-0.03$ & $-0.06$ & $-0.46$ \\
\addlinespace\midrule
\multicolumn{7}{l}{\textit{SD ratio --- model SD $/$ real SD; $1.0$ matches real spread}} \\
Cerebral cortex & $1.41$ & $0.95$ & $0.95$ & $0.99$ & $0.81$ & $0.78$ \\
Brain stem & $2.94$ & $1.10$ & $1.12$ & $1.04$ & $0.79$ & $0.85$ \\
Ventricles & $1.01$ & $1.03$ & $1.07$ & $1.19$ & $0.73$ & $0.71$ \\
Thalamus & $2.45$ & $1.03$ & $1.01$ & $0.97$ & $0.77$ & $0.86$ \\
Putamen & $2.55$ & $0.94$ & $0.94$ & $0.96$ & $0.77$ & $0.82$ \\
Cerebellar cortex & $2.91$ & $0.83$ & $0.88$ & $0.88$ & $0.76$ & $0.80$ \\
\bottomrule
\end{tabular}
\label{tab:drift}
\end{table}

Clean prediction and velocity prediction over-produce tissue, drifting positive on five and six of six structures respectively.
The flow-based strategies match the real means closely: flow matching, rectified flow, and the latent model are all within roughly $1\%$ of real on the thalamus, putamen, and cerebellar cortex.
Rectified flow and the latent model do so while generating less anatomical volume variation than real data, with a standard deviation ratio below $0.9$ on all six structures and as low as $0.71$.
Flow matching does not share this deficit, reproducing the real spread to within $0.88$ to $1.19$ across the six structures.

Table~\ref{tab:bootstrap} reports, for each strategy and structure, the percentage of draws in which the synthetic and real volumes could \emph{not} be distinguished; a higher percentage means the synthetic volumes were more often indistinguishable from real ones for that structure.
Flow matching was the strongest strategy overall, remaining indistinguishable from real data on the majority of draws for the thalamus~($85.1\%$), putamen~($85.4\%$), and cerebellar cortex~($90.6\%$).
Clean prediction achieved the single highest value in the table, $92.9\%$ on the putamen, but was distinguishable on every draw for four of the six structures.
Noise prediction and velocity prediction were distinguishable from real data on every draw for all six structures.
The cerebral cortex and the ventricles were not matched by any strategy, with no model exceeding $8.2\%$ of draws on either structure.

The three distributional measures we report~(FID, WASABI, and the per-structure Kolmogorov--Smirnov~(KS) test) do not rank the strategies identically.
The latent rectified-flow model achieves the best WASABI score in Table~\ref{tab:wasabi} yet is distinguishable from real on essentially every draw for five of six structures in Table~\ref{tab:bootstrap}; rectified flow shows a similar pattern.
The per-structure drift reported in Table~\ref{tab:drift} supports this: both LDM and rectified-flow strategies have accurate means, but too-narrow variances.
Flow matching is the only strategy with both accurate mean and variance, making it the strongest strategy of the per-structure test while placing second on the joint distance.

\begin{sidewaystable}[!ht]
\centering
\setlength{\tabcolsep}{6pt}
\caption{Permutation test on SynthSeg volumes. Each cell is the percentage of resampling draws in which a two-sample Kolmogorov--Smirnov test could \emph{not} distinguish that strategy's segmented volumes from real ones, at $p<0.05$. A higher percentage means the two samples were more often indistinguishable, i.e.\ the synthetic volumes are anatomically plausible for that structure. There are far more real test volumes ($8016$) than synthetic ones, and KS power grows with sample size, so each draw subsamples the real pool to $1000$ volumes --- the smallest synthetic count over the strategies compared --- and, where a synthetic pool is larger, subsamples it to the same size. Every cell therefore uses identical sample sizes, so the percentages are comparable across rows. $1000$ draws per cell, seed $0$. The percentage is over draws, not over image volumes: each draw compares two whole distributions and yields one $p$-value. Volumes are raw, not normalized by intracranial volume. All synthetic volumes were generated at the $0256$-step operating point. Highest in the table is clean prediction on putamen at $92.9\%$.}
\begin{tabular}{lrrrrrr}
\toprule
 & \textbf{Cerebral Cortex} & \textbf{Brain Stem} & \textbf{Ventricles} & \textbf{Thalamus} & \textbf{Putamen} & \textbf{Cerebellar Cortex} \\
\midrule
Noise ($\epsilon$) & 0.0\% & 0.0\% & 0.0\% & 0.0\% & 0.0\% & 0.0\% \\
Clean ($x_0$) & 0.0\% & 0.0\% & 0.0\% & 0.0\% & \textbf{92.9\%} & 23.6\% \\
Velocity & 0.0\% & 0.0\% & 0.0\% & 0.0\% & 0.0\% & 0.0\% \\
Flow & 0.0\% & 0.0\% & 0.0\% & \textbf{85.1\%} & 85.4\% & \textbf{90.6\%} \\
Rectified flow & \textbf{8.2\%} & \textbf{11.6\%} & 0.0\% & 0.1\% & 0.1\% & 2.0\% \\
LDM (rect.\ flow) & 0.1\% & 6.4\% & 0.0\% & 58.4\% & 0.0\% & 0.0\% \\
\bottomrule
\end{tabular}
\label{tab:bootstrap}
\end{sidewaystable}

\subsection{Observer Study Results}

Results are reported in Table~\ref{tab:observer}.
First, true real scans were read as real $79.5\%$ of the time~($737$ reads), providing an empirical ceiling for the study.
Across the $1{,}834$ reads that carry a true real-or-generated label~(real images and unconditional samples), raters identified the source correctly $64.5\%$ of the time.
On the samples, rectified flow was the most convincing at $66.5\%$ judged real, followed closely by the latent rectified-flow model at $64.0\%$ and flow matching at $57.2\%$.
No strategy reached the real-image ceiling.
On reconstructions, flow matching was highest at $72.6\%$, with clean prediction at $70.5\%$, while rectified flow fell to $35.2\%$.
The two strongest reconstruction rates approach the $79.5\%$ ceiling, which is a stronger result than any unconditional sample achieved and is consistent with the reconstruction quality reported in Table~\ref{tab:inverse}.

\begin{table}[!ht]
\centering
\small
\caption{Blind observer study. Percentage of lightboxes judged real by human raters who were not told which method produced any image, with Wilson 95\% confidence intervals. 2949 reads from 7 independent raters, none of them authors. Raters saw one lightbox at a time and judged it real or generated; an \emph{unsure} response is counted as \emph{not} real throughout, so a lightbox counts as passing for real only when a rater affirmatively said so. \emph{Generative samples} (Study A) are unconditional samples, where a higher rate means the sample passed for real. \emph{Reconstructions} (Study B) are solutions to the inverse problems; these have \emph{no} correct real/generated label --- a reconstruction of a real subject is neither --- so they are reported only as the rate judged real and never as an accuracy. \emph{Real images} is the study's ceiling. Bold marks the highest rate within a column.}
\begin{tabular}{lcc}
\toprule
& \multicolumn{2}{c}{\textbf{Judged real (\%, [95\% CI])}} \\
\cmidrule(lr){2-3}
\textbf{Strategy} & \textbf{Generative samples} & \textbf{Reconstructions} \\
& \textit{Study A} & \textit{Study B} \\
\midrule
\textit{Real images} & 79.5 \;[76.4, 82.3] & --- \\
\midrule
Noise ($\epsilon$) & 4.2 \;[2.0, 8.3] & 24.4 \;[19.0, 30.7] \\
Clean ($x_0$) & 31.0 \;[24.2, 38.6] & 70.5 \;[63.7, 76.6] \\
Velocity & 40.6 \;[33.7, 47.9] & 46.2 \;[38.9, 53.7] \\
Flow & 57.2 \;[50.2, 64.0] & \textbf{72.6} \;[65.9, 78.5] \\
Rectified flow & \textbf{66.5} \;[59.6, 72.7] & 35.2 \;[28.3, 42.7] \\
LDM (rect.\ flow) & 64.0 \;[57.2, 70.3] & 57.7 \;[50.7, 64.5] \\
\bottomrule
\end{tabular}
\label{tab:observer}
\end{table}

\subsection{Memory and time}
Eight NVIDIA H200 GPUs were used in parallel to achieve a training batch size of 24, each requiring 132 GiB of vRAM with a batch size of three.
Training took 82 minutes per epoch with 32-bit floating point precision.
We trained for 100 epochs, so the total training time was 8200 minutes, or 136.7 hours per model.

For inference, the models required 9 GiB of vRAM to generate a single sample.
To analyze the real-world time cost of generating samples, we performed inference with several NVIDIA GPUs with differing numbers of DDIM steps.
This is reported in Table~\ref{tab:gpu-times}.

\begin{sidewaystable}[!ht]
\centering
\setlength{\tabcolsep}{6pt} 
\caption{Inference time~(denoted as minutes:seconds) for a single sample from any model for different NVIDIA GPUs with differing numbers of DDIM steps~(Steps). The same times are true for all models. The inference time is roughly linear for the number of DDIM steps as well as for increased batch size.}
\begin{tabular}{rrrrrrrrr}
\toprule
\textbf{Steps} & \textbf{Tesla M40} & \textbf{A16} & \textbf{T4} & \textbf{RTX 8000} & \textbf{Tesla V100} & \textbf{A100} & \textbf{RTX 6000 ADA} & \textbf{H100 / H200} \\
\midrule
16   &  1:32 &  0:29 &  0:19 &  0:09 &  0:08 &  0:08 &  0:06 &  0:06 \\
32   &  3:07 &  0:59 &  0:39 &  0:18 &  0:18 &  0:17 &  0:12 &  0:11 \\
64   &  6:16 &  1:58 &  1:16 &  0:37 &  0:36 &  0:34 &  0:24 &  0:23 \\
128  & 12:24 &  3:57 &  2:31 &  1:15 &  1:12 &  1:08 &  0:49 &  0:47 \\
256  & 24:42 &  7:56 &  5:10 &  2:31 &  2:24 &  2:17 &  1:39 &  1:33 \\
512  & 49:46 & 15:54 & 10:25 &  5:02 &  4:48 &  4:34 &  3:19 &  3:07 \\
1000 & 96:24 & 31:03 & 20:22 & 12:27 &  9:23 &  8:56 &  6:30 &  6:06 \\
\bottomrule
\end{tabular}
\label{tab:gpu-times}
\end{sidewaystable}

\section{Discussion}
In this study, we presented MedForj for generating minimally preprocessed, high-resolution $1~\mathrm{mm}^3$ 3D $T_1$-w MR brain volumes.
While generative models exist for 3D human brain MRI, ours are the first to be trained on data of this size and variety, at a high isotropic resolution, without skullstripping, registration, or bias-field correction.
Despite its prevalence in applied DDPMs, we found clean prediction to yield less realistic results overall.
We also found velocity prediction, often treated as interchangeable with flow matching, to be substantially worse than flow matching on every measure we applied, including a mean drift of $+5\%$ to $+31\%$ on every segmented structure.
The flow-based strategies were the strongest of the six: flow matching gave the best inverse problem reconstructions, the highest rate of reconstructions judged real by blind raters, and the closest per-structure match to real segmented anatomy, while rectified flow and the latent rectified-flow model gave the best FID and the most convincing unconditional samples.

There are many potential applications of our model~\cite{koetzier2024generating,khosravi2025exploring}.
As a prior, MedForj can be used in routine image restoration tasks that benefit from a deep generative prior~\cite{chung2022diffusion,wang2023inversesr,rout2023solving,song2023solving}, such as super-resolution, denoising, and inpainting. 
We demonstrate this in our inverse problem solving experiment.
Additionally, it can be used for synthetic data generation, and since our model does not rely on pre-processing, the synthetic images can be used to train other models~\cite{frid2018gan,koetzier2024generating,fernandez2022can,akbar2024brain} in tasks such as bias field correction, registration~\cite{hoffmann2021synthmorph}, and skullstripping~\cite{hoopes2022synthstrip}.
Finally, in-line with current trends in foundation models~\cite{han2024parameterefficient,baklouti2025regularized,zhu2025new,d2025foundation}, our model may be used alongside fine-tuning for specific tasks for which there is insufficient training data.

In this work, we deliberately excluded images with gross focal pathology from the training corpus.
This choice was made to restrict what the prior is able to express.
The solution to an inverse problem balances between the data-consistency term and the prior term, and a choice along this balance has implications for pathology.
Pathology is preserved by data consistency whether or not the prior has ever seen a lesion, so a normative prior does not erase pathology; this is the regime of all five inverse problems reported here, each of which carries an explicit data-consistency term~\cite{chung2022diffusion,song2023solving}.
However, a prior that has the capability to generate pathology may do so even if no pathology exists in the data; for instance, in image inpainting, a pathological prior may generate a lesion but a normative prior would not.
In this work, we consider the possibility of a ``false positive'' to be more dangerous than a ``false negative.''
We did not measure lesion retention directly, and a specialized reconstruction study on pathological data is left as future work.

There are also limitations to our work.
First, the realism of our unconditional samples is empirically bounded.
Raters called genuine scans real $79.5\%$ of the time, whereas the most convincing unconditional samples~(from rectified flow) were called real only $66.5\%$ of the time, and raters identified image provenance correctly $64.5\%$ of the time, well above the $50\%$ chance rate~(Table~\ref{tab:observer}).
The distributional measures agree: the closest synthetic cohort still sits at $6.4$ times the real--real distance~(Table~\ref{tab:wasabi}), and no strategy matched the real cerebral cortex or ventricular volume distributions~(Table~\ref{tab:bootstrap}).
The story is different for inverse problem reconstructions: flow matching reconstructions were judged real $72.6\%$ of the time, approaching our real-image ceiling.
These models can be useful as \emph{priors} for recovering real anatomy.
Second, training these models is expensive.
The “pay-as-you-go” cost for an 8xH200 node on Microsoft Azure is \$84.80 per hour at the time of writing.
Training a single model for 136.7 hours costs approximately \$11,592; training all six models totals roughly \$69,552.
However, achieving the results in this paper required multiple trials and hyperparameter tuning. 
We estimated a total of 25 experiments, each averaging around 200 hours, leading to a naive minimum experimental cost of \$420,000.
However, at this level of compute use, alternative pricing options become more practical, such as buying hardware or reserving cloud instances~(which we used). 
These costs are high and the necessary hardware is not widely accessible, which motivates us to openly share our trained models with the broader community. 
Even with frozen, trained weights, our models require 9 GiB of vRAM when using automatic mixed precision at inference time. 
While consumer-grade GPUs typically have this amount of vRAM, it is still not common.
Finally, our models only generate structural $T_1$-w human brain images.
In practice, this is the most common contrast to acquire at high resolution~(indeed, this is how we were able to train our generative model in the first place).
Future work will fine-tune and adapt these models for additional contrasts and anatomies.

In conclusion, we trained six 3D diffusion models to generate high-resolution volumetric human brain MR images.
We found that the flow-based strategies yield improved results over noise, clean, and velocity prediction, and we recommend flow matching as the default MedForj prior.
We publicly released all models and code at~\url{https://github.com/piksl-research/medforj}.


\section{Materials and methods}
\label{sec:methods}
All data used in this study were retrospectively accessed under appropriate license and data use conditions.
See the supplementary material for data source information.

\subsection{Datasets and pre-processing}
\label{sec:datasets}
This study used data aggregated from 38 online datasets, named and shown in Table~\ref{tab:datasets}.
Data were omitted from the aggregated dataset if any of the following criteria were not met: 1) $T_1$-w; 2) 3D-acquisition; and 3) $\leq 1.2$mm resolution in all directions.
After aggregation, images were oriented to the axial orientation, interpolated with bicubic interpolation to $1$mm isotropic resolution, clipped to the original intensity range, padded or cropped to a size of $192\times 224\times 192$ voxels, then finally linearly min-max normalized and quantized to 16-bit integer values.
The pad and crop operation was performed with respect to the brain center of mass, found from the mask computed by HD-BET~\cite{Isensee-hdbet}.
We consider these operations to be ``minimal preprocessing'' because they do not reduce the visual variability of the data.

Some datasets contained multiple scans of the same subject, indicated in Table~\ref{tab:datasets}.
To avoid data leakage, our training and testing splits were done at the subject level.
Approximately 10\% of subjects were withheld from each dataset as testing data.
Additionally, two datasets were entirely withheld~(AIBL and SLEEP).
This yielded three distinct datasets: training data~(used to train the diffusion models), internal test data~(subjects withheld from training but from the same datasets as the training data), and external test data~(datasets withheld from training data).

The images that can be sampled from the generative models are a direct result of the training corpus; correspondingly, we describe the training corpus in the supplementary material in more detail, including population and clinical composition, published age range, country of acquisition, and population type.
The corpus is not a healthy-only sample.
Weighting each source by the number of training volumes it contributes, $67.7\%$ of the training data comes from cohorts recruited for aging, dementia, or movement-disorder research, so atrophy, ventricular enlargement, and other age- and disease-related variations are well represented; the curation protocol of Section~\ref{sec:curation} removed gross focal pathology and unacceptable image quality, not the diffuse chronic change that characterizes these cohorts.
The corpus includes subjects of age $5$ to $90+$ years and spans five continents, though not uniformly: pediatric and youth sources contribute only $2.3\%$ of training volumes.
The two withheld external sources are also distinct from each other and from the training corpus: AIBL is an Australian aging and Alzheimer's disease cohort~($\geq 60$ years), whereas SLEEP is a Polish study of healthy young adults~($18$--$35$ years).

\subsection{Curation protocol}
\label{sec:curation}
\begin{figure}[ht]
  \centering
  \begin{subfigure}[b]{0.48\textwidth}
    \centering 
    \includegraphics[width=\linewidth,page=2,trim=0 0 100 0,clip]{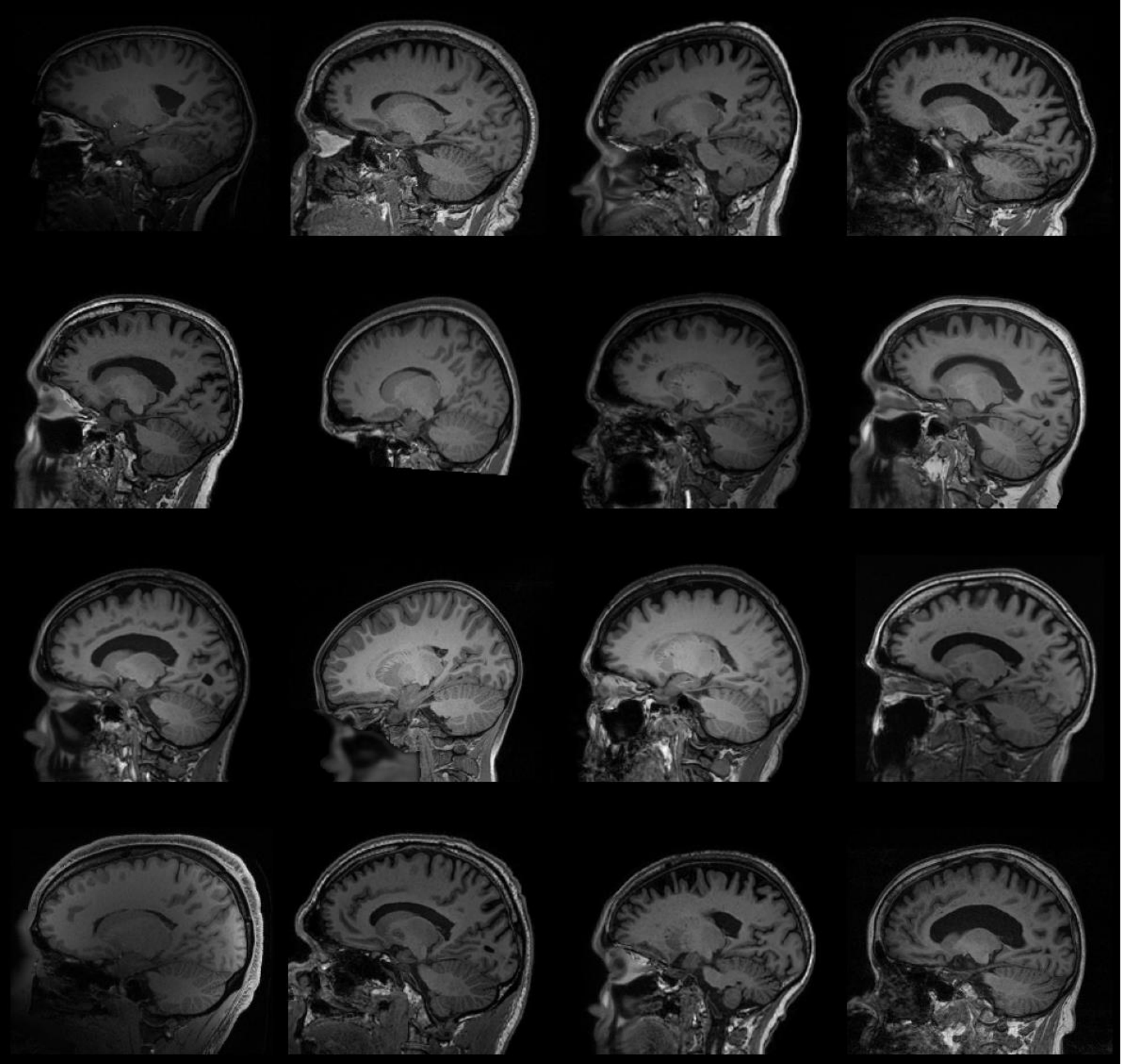}
    \caption{}
    \label{fig:repr-pass}
  \end{subfigure}
  \hspace{0.5em}
  \begin{subfigure}[b]{0.48\textwidth}
    \centering 
    \includegraphics[width=\linewidth,page=3,trim=0 0 100 0,clip]{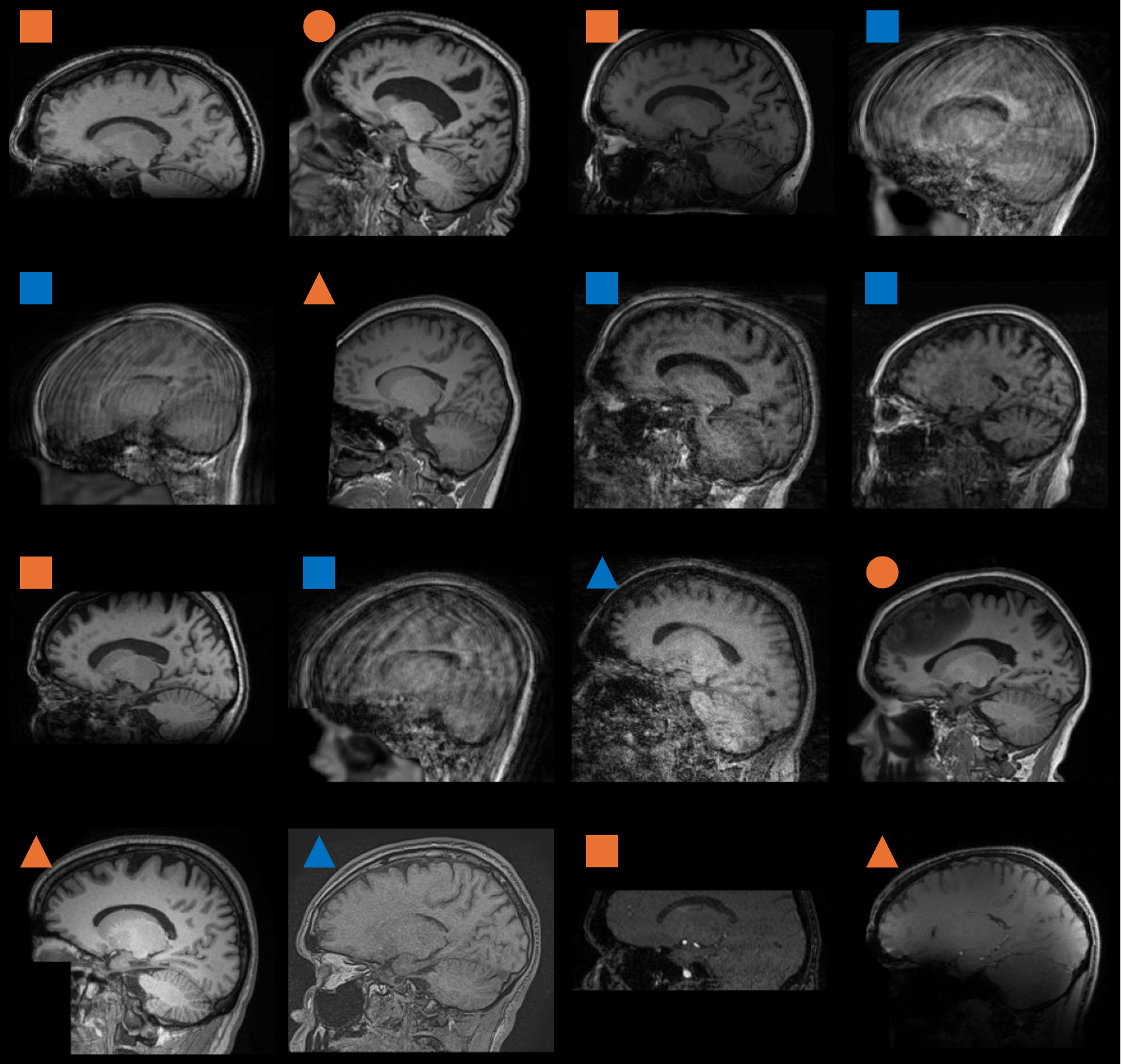}
    \caption{}
    \label{fig:repr-fail}
  \end{subfigure}
  \caption{Parasagittal slices from representative pass and fail images.~(a)~Images that passed manual quality assurance.~(b)~Images that failed manual quality assurance. Failure reasons include
      \textbf{Pathology}~(\textcolor{orange}{\scalebox{1.5}{$\bullet$}}),
      \textbf{Limited field-of-view~(FOV)}~(\textcolor{orange}{\scalebox{1.0}{$\blacksquare$}}),
      \textbf{Aggressive defacing}~(\textcolor{orange}{\scalebox{1.5}{$\blacktriangle$}}),
      \textbf{Motion artifacts}~(\textcolor{blue}{\scalebox{1.0}{$\blacksquare$}}),
      and \textbf{Noise}~(\textcolor{blue}{\scalebox{1.5}{$\blacktriangle$}}).}
  \label{fig:repr-pass-fail}
\end{figure}

After pre-processing, data underwent human quality assurance to exclude images with excessive lesions, noise, motion artifacts, limited field-of-view~(FOV), or other issues absent from the imaging header. 
Three expert raters~(S.W.R., A.C., B.E.D.) viewed every image and excluded those that met these criteria. This process was heuristic and leveraged the raters’ intuition and experience. Representative pass and fail images are shown in Figure~\ref{fig:repr-pass-fail}.

\subsection{Models}
\label{sec:model}
Generating synthetic images from a diffusion model involves progressively denoising an image over a series of steps.
A time-conditioned deep neural network, often using a U-net architecture~\cite{ddpm}, is trained to reverse a diffusion process that gradually corrupts data with Gaussian noise.
This process is modeled as a Markov chain that interpolates between the data distribution and an isotropic noise distribution.
We denote a clean data sample by $x_0$, a noise sample by $x_1$, and intermediate states by $x_t$ for $t\in[0, 1]$, where smaller values of $t$ correspond to lower noise levels.
The forward process, which describes how data diffuses into noise, has a closed-form solution~\cite{ddpm}, while the reverse process is learned and parameterized by a neural network.
Training aims to optimize parameters $\theta$ such that the neural network $f_\theta(x_t, t)$ estimates a less-noisy state $x_s$, for $s < t$.
During inference, a noise sample $x_1$ is drawn and successively denoised until the data state $x_0$ is obtained.

In the forward diffusion process, each state $x_t$ is expressed as a linear combination between data and noise:
\begin{equation}
  x_t = \alpha_t x_0 + \beta_t \epsilon,
\end{equation}
where $\alpha_t$ and $\beta_t$ are determined by the chosen noise schedule and $\epsilon \sim \mathcal{N}(\mathbf{0}, \mathbf{I})$ by definition.
In DDPMs as well as this paper, $\alpha_t^2 + \beta_t^2 = 1$ for all $t$, ensuring variance preservation across the entire diffusion trajectory.

The reverse process is defined similarly, but in terms of the estimated signal and noise:
\begin{equation}
  x_{s} = \alpha_{s} \hat{x}_0 + \beta_{s} \hat{\epsilon},
\label{eq:ddpm-ddim}
\end{equation}
where $\hat{\epsilon} = (x_t - \alpha_t \hat{x}_0) / \beta_t$.
Denoising proceeds by first estimating $\hat{x}_0$ from time $t$, then projecting this estimate to an earlier time point $s$~(where $s < t$) along the diffusion path.
This formulation follows DDIM~\cite{DDIM}, which extends the DDPM framework~\cite{ddpm} by enabling deterministic and computationally efficient sampling.

During inference, sampling with DDIM requires the pre-defined noise schedule parameters $\alpha_{s}$ and $\beta_{s}$, an initial sample $x_1 \sim \mathcal{N}(\mathbf{0}, \mathbf{I})$, and a trained time-conditioned function $f_\theta(\cdot, t)$.
Given the update rule in Eq.~\ref{eq:ddpm-ddim}, reconstructing the clean image $\hat{x}_0$ requires $\hat{\epsilon}$~(or vice-versa).
Accordingly, the role of $f_\theta$ is to estimate one of these quantities from the noisy input at time $t$.
Prior work~\cite{gao2025diffusionmeetsflow,jit} has identified four standard parameterizations for the network output.

In \textit{noise prediction}, the network $f_\theta$ directly estimates $\hat{\epsilon} = f_\theta(x_t, t)$.
Then
\begin{equation}
  \hat{x}_0 = (x_t - \beta_t f_\theta(x_t, t)) / \alpha_t.
\end{equation}

In \textit{clean prediction}, the network $f_\theta$ estimates $\hat{x}_0 = f_\theta(x_t, t)$.
Then,
\begin{equation}
  \hat{\epsilon} = (x_t - \alpha_t f_\theta(x_t, t)) / \beta_t.
\end{equation}

In \textit{velocity prediction}, rather than directly estimating $\hat{x}_0$ or $\hat{\epsilon}$, the network $f_\theta$ estimates the instantaneous velocity field at time $t$ moving towards $x_0$~\cite{kingma2021variational,kingma2023understanding}: $\hat{v} = \alpha_t \hat{\epsilon} - \beta_t \hat{x}$.
Then, $\hat{v} = f_\theta(x_t, t)$ and 
\begin{equation}
  \hat{x} = (\alpha_t \hat{\epsilon} - f_\theta(x_t, t)) / \beta_t,
\end{equation}
and
\begin{equation}
  \hat{\epsilon} = (f_\theta(x_t, t) + \beta_t \hat{x}) / \alpha_t.
\end{equation}

Finally, in \textit{flow matching}, the network $f_\theta$ estimates the flow-matching field: $\hat{u} = \hat{\epsilon} - \hat{x} = f_\theta(x_t, t)$, which yields
\begin{equation}
  \hat{x} = \hat{\epsilon} - f_\theta(x_t, t)
\end{equation}
and
\begin{equation}
  \hat{\epsilon} = f_\theta(x_t, t) + \hat{x}.
\end{equation}

For some choices of $\alpha_t$ and $\beta_t$, the velocity and flow formulations are identical~\cite{jit}.
However, with variance preservation they are not~\cite{gao2025diffusionmeetsflow}.
\textit{Rectified flow} is a variant of the flow matching formulation, but sets $\alpha_t$ and $\beta_t$ such that all states $x_t$ are convex combinations of $x_0$ and $\epsilon$. 

Each of the four prediction targets also yields a corresponding loss function.
With algebraic manipulation, it is possible for the $f_\theta$ to produce one prediction target while computing a loss in a different prediction target~\cite{jit}; with four prediction targets, there are 16 total possible choices.
However, in this work, we investigated five and accordingly trained five voxel-space diffusion models, all with a common software framework~\cite{MONAI}.
These models used noise, clean, velocity, and flow prediction targets and losses.
The fifth model used clean prediction with velocity loss as proposed by Li and He~\cite{jit}.
A sixth model applied this same rectified-flow strategy in the latent space of the pretrained MAISI autoencoder~\cite{guo2025maisi,zhao2026maisi}, giving the six training strategies compared in this work.

Each diffusion model used the same U-Net~\cite{ronneberger2015unet} backbone: five levels with 16, 32, 64, 128, and 256 channels respectively. 
We trained using the conventional $T=1000$ diffusion steps~\cite{ddpm}.
The number of DDIM~\cite{DDIM} steps used at inference was selected empirically as $256$, following the ablation reported in Table~\ref{tab:ablation-steps}.
We used the Adam optimizer~\cite{kingma2015adam} with a learning rate of $10^{-4}$ and batch size of 24, implemented as a batch of three on each of eight NVIDIA H200 GPUs.
We trained for $100$ epochs with random 3D rotations~(with linear interpolation) of up to $\pm10^\circ$ and 3D translations of up to $\pm5$ mm.

\subsection{Inverse problem solving}
\label{sec:inverse-methods}
Each of these inverse problems was implemented over $100$ subjects with simulations.
The five voxel-space priors were solved with the diffusion posterior sampling~(DPS)~\cite{chung2022diffusion} family of techniques.
The latent model used ReSample~\cite{song2023solving}, an adaptation of DPS for LDMs.

The $100$ subjects were held out from training and drawn evenly from the two external test sites~($50$ from AIBL~\cite{dataset-aibl} and $50$ from SLEEP~\cite{dataset-sleep}).
Every volume was $192 \times 224 \times 192$ at $1$~mm isotropic resolution and was percentile-normalized to $[-1, 1]$ using the $0.5$th and $99.5$th percentiles as the intensity bounds, matching the convention the models were trained under.
Writing $x$ for the ground truth volume and $\mathcal{A}$ for a forward operator, each measurement $y = \mathcal{A}(x)$ was simulated once, written to disk in floating point so that the $[-1, 1]$ scale was preserved exactly, and then reused verbatim by every strategy.

\emph{Through-plane super-resolution} was modeled as slice-selective excitation along the superior-inferior axis with a slice thickness of $5$~mm and slice separation of $6.5$~mm~(therefore a gap of $1.5$~mm).
The observation matrix $A \in \mathbb{R}^{29 \times 192}$ has rows given by the slice profile
\begin{equation}
A_{ij} \;=\; \frac{\Delta}{\sqrt{2\pi}\,\sigma} \exp\!\left(-\frac{(z_i - z_j)^2}{2\sigma^2}\right),
\end{equation}
where $z_j$ are the $192$ high-resolution slice positions at $\Delta = 1$~mm spacing, $z_i$ are the low-resolution slice positions, and $\sigma$ is obtained by deconvolving the high-resolution sampling from the slice thickness, $\sigma^2 = \sigma_{5\,\mathrm{mm}}^2 - \sigma_{1\,\mathrm{mm}}^2$ with $\sigma_\gamma = \gamma / (2\sqrt{2\ln 2})$, giving $\sigma = 2.08$~mm for a $5$~mm slice thickness.
Slices were placed at a separation of $6.5$~mm, so $y = Ax$ has $29$ slices in place of $192$.
This operator is linear and its adjoint $A^\top$ is available in closed form.

\emph{Inpainting} removed a fixed interior block, $y = m \odot x$, where $\odot$ denotes the Hadamard product, and the mask $m$ is zero on the central $64 \times 96 \times 64$ box and one elsewhere.
The operator is linear and self-adjoint~\cite{ddnm}.

\emph{Rician denoising} used the magnitude noise model appropriate for MRI~\cite{gudbjartsson1995rician} rather than additive Gaussian noise.
Denoting a $[0, 1]$ intensity normalization as $x_{01} = (x+1)/2$, the measurement is
\begin{equation}
y_{01} \;=\; \sqrt{(x_{01} + n_1)^2 + n_2^2}, \qquad n_1, n_2 \sim \mathcal{N}(0, \sigma_n^2),
\end{equation}
with $\sigma_n = 0.1$ in $[0,1]$ intensity units, renormalized to $[-1, 1]$ as $y = 2y_{01} - 1$.
When solving, the deterministic part of this model is the identity, so the forward operator passed to the solver is the identity and the Rician character of the corruption is carried entirely by the data-consistency likelihood described below.

\emph{k-space acceleration} simulated an undersampled 3D Cartesian acquisition,
\begin{equation}
y \;=\; \left| \mathcal{F}^{-1} \!\left( M \odot \mathcal{F} x_{01} \right) \right|,
\end{equation}
where $\mathcal{F}$ is the centered, orthonormal 3D discrete Fourier transform and $M$ is a phase-encode mask defined over the two phase-encode axes~($192 \times 224$) and broadcast along the readout axis.
The mask is variable-density: the central $8\%$ of each phase-encode axis~(a $15 \times 18$ block of low frequencies) is fully sampled, and the remaining phase-encode locations are drawn without replacement with probability proportional to $(1-r)^4$, where $r$ is the normalized radius from the k-space origin.
Lines were drawn until the total reached $1/R$ of the phase-encode plane at $R = 12$, i.e.\ $3{,}584$ of $43{,}008$ locations.
No density compensation or energy rescaling was applied, so the zero-filled reconstruction loses energy as well as detail.
A single fixed mask realization was shared by every subject and by the solver.
Because the operator returns a magnitude image after the inverse transform, it is nonlinear and has no adjoint.

\emph{Motion correction} used a k-space-domain rigid motion simulation in the style of TorchIO's \texttt{RandomMotion}~\cite{perez2021torchio}.
A single motion event occurs at $45\%$ through the acquisition, displacing the head by $6^\circ$ about each of the three axes and $6$~mm along each of the three axes.
The stationary object and the displaced object are each transformed to k-space, and the readout axis is partitioned into contiguous slabs whose boundaries fall at the motion event time: lines $0$ to $86$ are taken from the spectrum of the displaced pose and lines $86$ to $192$ from the stationary pose.
Following the TorchIO convention, the spectra are ordered so that the slab containing the k-space origin carries the stationary spectrum, which keeps the gross contrast of the volume intact while corrupting the high-frequency structure into the ghosting and blurring characteristic of in-plane motion.
The real part of the inverse transform is taken as the measurement.
This operator resamples the object and returns a real image, so it is nonlinear and has no adjoint.

For the five voxel-space priors, solving proceeded by DPS over $256$ steps, matching the sampling operating point selected in Section~\ref{sec:model}.
At each step the clean estimate $\hat{x}_{0|t}$ was formed from the network output according to that model's parameterization, clamped to $[-1, 1]$, and the noise estimate recomputed from the clamped $\hat{x}_{0|t}$ so that the two remain a consistent decomposition of $x_t$.
The measurement then enters as a gradient on the noisy state,
\begin{equation}
x_{s} \;=\; \sqrt{\bar{\alpha}_s}\, \hat{x}_{0|t} + \sqrt{1 - \bar{\alpha}_s - \sigma_t^2}\, \hat{\epsilon}_t \;-\; \zeta \nabla_{x_t} \ell\!\left( \mathcal{A}(\hat{x}_{0|t}),\, y \right) \;+\; \sigma_t z,
\end{equation}
with $z \sim \mathcal{N}(0, I)$ and $\eta = 1$.
The data term $\ell$ is the $\ell_2$ norm $\lVert \mathcal{A}(\hat{x}_{0|t}) - y \rVert_2$ for the Gaussian case, and for Rician denoising it is the Rician negative log likelihood, which after dropping terms independent of the reconstruction is
\begin{equation}
\ell \;=\; \sum \left[ \frac{\hat{x}^2}{2\sigma_n^2} - \log I_0\!\left( \frac{y\,\hat{x}}{\sigma_n^2} \right) \right],
\end{equation}
evaluated in $[0,1]$ units, where $I_0$ is the zeroth-order modified Bessel function of the first kind, evaluated in the log domain for numerical stability.
Rectified flow is the exception to the single-gradient update: because a single DPS-style correction is unstable under the rectified-flow parameterization, its likelihood correction is applied as an inner loop of $50$ gradient steps on the clean estimate, which is then blended with the uncorrected estimate as $(1-\tau)\hat{x}_0 + \tau x_0^{\mathrm{corr}}$ so that the correction is emphasized early in the trajectory and relaxed near the end, and the noise component is stochastically renoised.

The guidance weight $\zeta$ was calibrated per parameterization, because the scale of the gradient differs substantially between them.
Gaussian tasks used $\zeta = 500$ for epsilon and velocity prediction and $\zeta = 20$ for clean prediction, flow matching, and rectified flow.
Rician denoising used $\zeta = 5 \times 10^{-4}$ for all voxel-space priors, selected by a five-point sweep over $\{5 \times 10^{-4}, 5\times 10^{-3}, 5 \times 10^{-2}, 5 \times 10^{-1}, 5\}$ and confirmed visually against the noisy measurement; the Rician likelihood carries a $1/\sigma_n^2$ factor and so sits at a very different scale from the Gaussian data term.

The latent model cannot take a voxel-space gradient directly, so it used ReSample~\cite{song2023solving}, which alternates unconditional latent steps with hard data consistency enforced in the image domain.
At $8$ evenly spaced events over the $256$ steps, the latent estimate $\hat{z}_0$ is decoded, corrected against the measurement, re-encoded, and stochastically resampled back onto the trajectory at time $\tau_s$ with variance $\gamma \tau_s^2$ for $\gamma = 40$.
For the two linear problems, super-resolution and inpainting, data consistency is solved exactly as the proximal problem $(A^\top A + \lambda I)\,x = A^\top y + \lambda x_{\mathrm{init}}$ by $20$ iterations of conjugate gradient with $\lambda = 10^{-2}$.
For the three nonlinear problems, k-space acceleration, motion, and Rician denoising, no adjoint exists and data consistency is instead minimized by gradient descent through the forward operator using the same measurement likelihoods as above.
The correction step was run in the same $[-1,1]$ convention as the voxel-space operators, bridging through the autoencoder, whose native range is $[0,1]$.
We fixed the number of resampling events rather than the interval between them, since holding an absolute step interval fixed while increasing the number of sampling steps silently multiplies the number of lossy encode-decode round-trips, which compound late in the trajectory.

Finally, a conventional no-prior ``baseline'' was included on every task so that the priors are measured against what the measurement already gives for free.
The baseline is the input measurement itself in image space with no generative model: trilinear interpolation back to $192 \times 224 \times 192$ for super-resolution, and $y$ unchanged for the other four tasks.

\subsection{Fr\'echet Inception Distance}
\label{sec:fid-methods}
Since FID~\cite{heusel2017gans} leverages a pre-trained feature extractor in 2D for natural images, FID was calculated on 2D slices separated by $4$~mm extracted from axial, coronal, and sagittal orientations.
The library \texttt{pytorch-fid}~\cite{Seitzer2020FID} was used to compute the FID.

\subsection{WASABI}
\label{sec:wasabi-methods}
To briefly summarize WASABI, each subject can be represented as a point whose coordinates are their segmented structure volumes~(segmentations computed by SynthSeg~\cite{billot_synthseg_2023}), normalized by total intracranial volume so that overall head size does not dominate.
A real and a synthetic cohort are then two clouds of such points, and WASABI reports the Wasserstein~(optimal transport) distance between them: the least total work needed to move one cloud onto the other, computed over all structures jointly.
Because the absolute scale of an optimal transport distance is not interpretable on its own, we also computed the same distance between two disjoint halves of the real data, which is the score a perfectly matched synthetic cohort would achieve.

\subsection{Permutation testing of segmentation distributions}
\label{sec:permutation-methods}
The Kolmogorov--Smirnov~(KS) test~\cite{massey1951kolmogorov} was used to assess whether the segmented volume distributions of the synthetic images differed from those of the real images.
Because there are roughly eight times more real test volumes than synthetic volumes~($8{,}016$ versus $1{,}000$; see Table~\ref{tab:datasets}) and the power of the KS test grows with sample size, a permutation procedure was used so that every comparison is made at an identical sample size.
Each of $1{,}000$ draws subsampled the real pool to $1{,}000$ volumes, ran the KS test, and recorded whether the two samples could be distinguished at $p<0.05$.

\subsection{Observer study}
\label{sec:observer-methods}
We ran a blind observer study with seven independent (non-author) raters.
Each viewed triplanar lightbox montages one at a time and gave a verdict of \emph{real}, \emph{generated}, or \emph{unsure}.
Raters were never told which method produced any image, and no identifying text appeared in any montage.
Five raters performed at least $100$ reads, and $2{,}949$ reads were collected in total.
Verdicts were scored conservatively: only an affirmative \emph{real} counts as judged real; an \emph{unsure} response was not promoted to \emph{real}.

Each reader was presented with a uniformly randomly sampled image from one of three pools of data.
\emph{Real data} are image volumes selected from the external test set.
\emph{Generative samples} are unconditional samples from each model, for which a correct answer exists and a higher judged-real rate means the sample passed for real.
\emph{Reconstructions} are solutions to the five inverse problems; these have no correct real-or-generated label, since a reconstruction of a real subject is neither, and are therefore reported only as the rate judged real and never as an accuracy.

\backmatter

\bmhead{Funding}
This work is supported in part by the DoD CDMRP Award HT94252410785~(PI:Dewey), National MS Society RG-2410-44033~(PI:Dewey), the NIH under NIBIB grant R01-EB036013~(PI: J.L.~Prince) and by the Congressionally Directed Medical Research Programs Grant W81XWH2010912~(PI: J.L.~Prince).
This material is partially supported by the National Science Foundation Graduate Research Fellowship under Grant No. DGE-1746891~(S.W.~Remedios).
GPU compute time was made possible through a grant from the Johns Hopkins Data Science and AI Institute and Johns Hopkins Research IT.

\bmhead{Data availability}
All datasets used are publicly available.
Their sources and appropriate data use statements are shown in Table~\ref{tab:datasets} and the supplementary material.

\bmhead{Code availability}
We released the trained model weights for all six training strategies~(noise prediction, clean prediction, velocity prediction, flow matching, rectified flow, and latent diffusion rectified flow) as well as the training and inference code.
The code and weights are available online at \url{https://github.com/piksl-research/medforj}.

\bmhead{Author contributions}
S.W.R. and B.E.D. designed the research.
S.W.R, A.C., and B.E.D. obtained and manually curated the data.
S.W.R. and B.E.D. implemented the experiments.
S.W.R. and B.E.D. analyzed and discussed the experimental results.
S.W.R. prepared the manuscript.
J.L.P. and B.E.D. supervised the project and ensured administrative support.
All authors contributed to reviewing and revising the manuscript and have approved the manuscript for final submission.

\bmhead{Competing interests}
The authors declare no competing interests.

\bibliography{sn-bibliography}

\input{consortia}

\end{document}

%% file: consortia.tex
%

\newcommand{\cbase}{2}
\newcommand{\caff}[1]{\the\numexpr\cbase+#1\relax}
\newcommand{\cs}[1]{\textsuperscript{\caff{#1}}}

\section*{Alzheimer's Disease Neuroimaging Initiative}

Data used in the preparation of this article were obtained from the
Alzheimer's Disease Neuroimaging Initiative (ADNI) database
(\url{adni.loni.usc.edu}). As such, the investigators within the ADNI
contributed to the design and implementation of ADNI and/or provided data
but did not participate in analysis or writing of this report. A complete
listing of ADNI investigators can be found at:
\url{http://adni.loni.usc.edu/wp-content/uploads/how_to_apply/ADNI_Acknowledgement_List.pdf}.

\section*{Australian Imaging Biomarkers and Lifestyle Flagship Study of Aging}

Data was partially collected by the AIBL study group. AIBL study methodology
has been reported previously (Ellis et al. 2009) Data used in the preparation
of this article was obtained from the Australian Imaging Biomarkers and
Lifestyle flagship study of ageing (AIBL) funded by the Commonwealth
Scientific and Industrial Research Organisation (CSIRO) which was made
available at the ADNI database (\url{www.loni.usc.edu/ADNI}). The AIBL
researchers contributed data but did not participate in analysis or writing
of this report. AIBL researchers are listed at \url{www.aibl.csiro.au}.

\section*{Health and Aging Brain Study (HABS-HD) Study Team}

HABS-HD MPIs: Sid E O'Bryant\cs{1}, Kristine Yaffe\cs{2}, Arthur Toga\cs{3},
Robert Rissman\cs{4}, \& Leigh Johnson\cs{1}; and the HABS-HD Investigators:
Meredith Braskie\cs{3}, Kevin King\cs{4}, James R Hall\cs{1},
Melissa Petersen\cs{1}, Raymond Palmer\cs{5}, Robert Barber\cs{1},
Yonggang Shi\cs{3}, Fan Zhang\cs{1}, Rajesh Nandy\cs{1},
Roderick McColl\cs{6}, David Mason\cs{1}, Bradley Christian\cs{7},
Nicole Philips\cs{1}, Stephanie Large\cs{1}, Joe Lee\cs{8},
Badri Vardarajan\cs{8}, Monica Rivera Mindt\cs{9}, Amrita Cheema\cs{10},
Lisa Barnes\cs{11}, Mark Mapstone\cs{12}, Annie Cohen\cs{13},
Amy Kind\cs{7}, Ozioma Okonkwo\cs{7}, Raul Vintimilla\cs{1},
Zhengyang Zhou\cs{1}, Michael Donohue\cs{4}, Rema Raman\cs{4},
Matthew Borzage\cs{4}, Michelle Mielke\cs{14}, Beau Ances\cs{15},
Ganesh Babulal\cs{15}, Jorge Llibre-Guerra\cs{15}, Carl Hill\cs{15} and
Rocky Vig\cs{16}. Research reported on this publication was supported by the
National Institute on Aging of the National Institutes of Health under Award
Numbers R01AG054073, R01AG058533, P41EB015922 and U19AG078109. The content is
solely the responsibility of the authors and does not necessarily represent
the official views of the National Institutes of Health.

\medskip
{\small
\noindent\cs{1}University of North Texas Health Science Center, Fort Worth, TX, USA.\\
\cs{2}UCSF Weill Institute for Neurosciences, University of California San Francisco, San Francisco, CA, USA.\\
\cs{3}Stevens Neuroimaging and Informatics Institute, Keck School of Medicine, University of Southern California, Los Angeles, CA, USA.\\
\cs{4}Keck School of Medicine, University of Southern California, Los Angeles, CA, USA.\\
\cs{5}The University of Texas Health Science Center at San Antonio, San Antonio, TX, USA.\\
\cs{6}The University of Texas Southwestern Medical Center, Dallas, TX, USA.\\
\cs{7}University of Wisconsin School of Medicine and Public Health, Madison, WI, USA.\\
\cs{8}Columbia University, New York, NY, USA.\\
\cs{9}Fordham University, New York, NY, USA.\\
\cs{10}Georgetown University, Washington, DC, USA.\\
\cs{11}Rush University Medical Center, Chicago, IL, USA.\\
\cs{12}University of California Irvine School of Medicine, Irvine, CA, USA.\\
\cs{13}University of Pittsburgh, Pittsburgh, PA, USA.\\
\cs{14}Wake Forest University School of Medicine, Winston-Salem, NC, USA.\\
\cs{15}Washington University in St. Louis, St. Louis, MO, USA.\\
\cs{16}Midtown Medical Imaging, Fort Worth, TX, USA.
}

\section*{Mayo Clinic Study of Aging}

Data used in preparation of this article were shared by the Mayo Clinic Study
of Aging (MCSA). The MCSA is funded by the following sources: NIH U01
AG006786, R01 AG034676, R37 AG011378, R01 AG041851, R01 NS097495, R01
AG056366, R01 AG068206, P30 AG062677, GHR Foundation, Elsie and Marvin
Dekelboum Family Foundation, Liston Award, Schuler Foundation, Alexander
Foundation, Mayo Foundation for Medical Education and Research.